\documentclass[%
 reprint,
 amsmath,amssymb,
 aps,
 pra,
]{revtex4-2}

\usepackage{graphicx}
\usepackage{dcolumn}
\usepackage{bm}
\usepackage{braket}
\usepackage{tikz}
\usepackage{quantikz}
\usepackage[hidelinks]{hyperref}
\usepackage{subcaption}
\usepackage{mathtools}
\usepackage{xcolor}

\usetikzlibrary{decorations.pathreplacing}

\begin{document}

\preprint{APS/123-QED}

\title{Quantum Algorithm for Solving the Advection Equation \\ using Hamiltonian Simulation}

\author{Peter Brearley}
 \email{p.brearley@imperial.ac.uk}
\author{Sylvain Laizet}%
\affiliation{%
Department of Aeronautics, Imperial College London, London, United Kingdom.
}%

\date{June 14, 2024}

\begin{abstract}
    A quantum algorithm for solving the advection equation by embedding the discrete time-marching operator into Hamiltonian simulations is presented. One-dimensional advection can be simulated directly since the central finite difference operator for first-order derivatives is anti-Hermitian. Here, this is extended to industrially relevant, multi-dimensional flows with realistic boundary conditions and arbitrary finite difference stencils. A single copy of the initial quantum state is required and the circuit depth grows linearly with the required number of time steps, the sparsity of the time-marching operator and the inverse of the allowable error. Statevector simulations of a scalar transported in a two-dimensional channel flow and lid-driven cavity configuration are presented as a proof of concept of the proposed approach. 
\end{abstract}

\maketitle


\section{Introduction}

Quantum computing is expected to bring a profound shift in our computational capability. Among the most promising applications is solving large-scale partial differential equations (PDEs) more efficiently than classical computers. PDEs are ubiquitous across science and engineering and solving them currently occupies the majority of the world's high-performance computing resources. The development of efficient quantum algorithms is therefore of immense value and has attracted a large interdisciplinary research community, spurred on by the continuing advancements in quantum hardware \citep{Bravyi2022, Stephenson2022}.

The advection equation is a foundational linear PDE spanning multiple industries as it describes the transport of a scalar quantity in advection-dominated flows. It is given by
\begin{equation}
    \frac{\partial \phi}{\partial t} + u_j\frac{\partial \phi}{\partial x_j} = 0
    \label{eq:advection_equation}
\end{equation}
where $\phi$ represents the scalar field (e.g.\ temperature, concentration) and $u_j$ is the $j$\textsuperscript{th} component of the advective velocity vector, where repeated indices invoke summation over all spatial dimensions. Applications include modelling the vast oceanic \cite{Webb1998}, atmospheric \cite{Rood1987}, and geological flows \cite{Samuel2010} used in climate studies, drug delivery systems in biomathematics \cite{Bazilevs2007}, and heat exchangers for cooling oil refineries, chemical processing plants and power stations \cite{Diao2004}.  When the advection equation is discretised in space using the finite difference method, the homogeneous ordinary differential equation (ODE)
\begin{equation}
    \frac{d\vec{\phi}}{d t} = M\vec{\phi}
    \label{eq:ODE}
\end{equation}
is produced. The original motivation for quantum computing was to simulate quantum dynamics governed by the Schr\"{o}dinger equation \cite{Feynman1982}, which can be expressed in the form of Eq.\ \eqref{eq:ODE} for an anti-Hermitian coefficient matrix $M$. The central finite difference operator for first-order derivatives is anti-Hermitian, so for one-dimensional flows with periodic boundary conditions, advection can be simulated on a quantum computer using the unitary operator $e^{Mt}$. This implies that one-dimensional advection can be considered equivalent to quantum dynamics \cite{An2022}. Indeed, advection in a divergence-free velocity field is inherently a norm-preserving process, making it well-suited for simulation on a quantum computer. However, for industrially relevant multidimensional flows with realistic boundary conditions or non-centred finite difference stencils, $M$ ceases to be anti-Hermitian so a different approach must be taken.

Quantum algorithms for solving PDEs can be delineated into two categories: fully quantum algorithms that implement quantum circuits to evolve the quantum state as described by the PDE of interest, and quantum-classical hybrid algorithms where a quantum computer is used for a specific task in a larger, classical computation. Fully quantum approaches generally excel at solving linear PDEs because quantum operators act linearly on quantum superpositions, allowing algorithms based on the finite difference method (FDM) \citep{Cao2013, Costa2019, Wang2020, Childs2021}, the finite element method (FEM) \citep{Clader2013, Montanaro2016} and spectral methods \citep{Childs2020} to be effectively represented quantum mechanically. Encoding the solution from $N=2^n$ grid points within the amplitudes $\alpha_j$ of an $n$-qubit quantum state $\ket{\psi} = \sum_{j=1}^{N} \alpha_j \ket{j}$ leads to an exponentially growing capacity to store information and an inherent quantum parallelism when processing it. Amplitude encoding does not allow for the inspection of the full solution as with classical methods, but rather the extraction of global statistics into the limited output space. This may be adequate depending on the context and make previously intractable problems tractable. Fully quantum approaches are not limited to linear PDEs as techniques have been proposed based on the derivation of the nonlinear Schr\"{o}dinger equation using mean-field techniques \citep{Lloyd2020} and Carleman linearisation \citep{Liu2021} to tackle nonlinear PDEs, but these are generally limited to weakly nonlinear interactions. On the other hand, variational quantum algorithms (VQAs) \citep{Peruzzo2014} for solving optimisation problems have been used as the basis for hybrid algorithms that have demonstrated a greater capability of tackling nonlinear PDEs \citep{Lubasch2020, Kyriienko2021, Jaksch2023}. \citeauthor{Kyriienko2021} \cite{Kyriienko2021} used a machine learning strategy where differentiable quantum circuits were trained to solve nonlinear differential equations. \citeauthor{Jaksch2023} \cite{Jaksch2023} extended a quantum algorithm for solving nonlinear problems \cite{Lubasch2020} to fluid dynamics, evaluating cost functions from matrix product state representations of the flow \citep{Gourianov2022} to obtain a polynomial upper bound on the depth of the variational network. VQA-based algorithms are of interest because of their potential to operate on near-term hardware, \citep{Peruzzo2014} though a definitive quantum advantage is yet to be demonstrated.

Most quantum algorithms for solving linear PDEs have a quantum linear systems algorithm (QLSA) at their core \citep{Clader2013, Cao2013, Berry2014, Montanaro2016, Berry2017, Arrazola2019, Wang2020, Childs2020, Childs2021, Krovi2022, Berry2022} such as the HHL algorithm \citep{Harrow2009} or further optimisations thereof \citep{Childs2017, Costa2022}. \citeauthor{Clader2013} \cite{Clader2013} developed a quantum algorithm using a QLSA to implement the FEM for solving Maxwell's equations, and this was further clarified and developed by \citeauthor{Montanaro2016} \cite{Montanaro2016}. \citeauthor{Cao2013} \cite{Cao2013}, \citeauthor{Wang2020} \cite{Wang2020} and \citeauthor{Childs2021} \cite{Childs2021} in their respective studies optimised a quantum algorithm based on the FDM to solve the Poisson equation by expressing the PDE as a system of linear equations and then solving with a QLSA. Algorithms for solving homogeneous, time-independent ODEs \cite{Berry2014, Berry2017, Krovi2022, Berry2022} in the form of Eq.\ \eqref{eq:ODE} can be applied to the spatially discretised advection equation. \citeauthor{Berry2014} \cite{Berry2014} proposed an algorithm using a QLSA with a linear multistep method, where the error per time step is a high power of the time step size. A different approach was later taken by \citeauthor{Berry2017} \cite{Berry2017} by encoding a truncated Taylor series expansion of $e^{Mt}$ in a linear system of equations, achieving an exponentially improved dependence on precision. This work was later extended by \citeauthor{Krovi2022} \cite{Krovi2022} to non-diagonalisable and singular matrices, while also achieving an exponential improvement over \citeauthor{Berry2017} \cite{Berry2017} for diagonalisable matrices with a bounded value of $||e^{Mt}||$. \citeauthor{Berry2022} \cite{Berry2022} proposed encoding a truncated Dyson series into a system of linear equations, achieving a scaling that is linear in the evolution time $T$ and the norm $||M||$, and poly-logarithmic in the allowable error $\epsilon$. The primary disadvantage of algorithms that employ a QLSA is that they require a large number of state initialisation queries that grow linearly with the condition number of the matrix \cite{Harrow2009}, which itself grows linearly with the desired simulation time \cite{An2023}. This introduces a potentially prohibitive computational overhead given the challenges of state preparation. In addition, the runtime of the HHL algorithm has a dependency on $1/\epsilon$ arising from the quantum phase estimation (QPE) step, which may become prohibitive for algorithms requiring repeated applications. Subsequent works \cite{Childs2017, Costa2022} have improved the $\epsilon$ dependence by applying the inverse of the matrix as a linear combination of unitaries (LCU) \cite{Childs2017} or by the quantum adiabatic theorem \cite{Costa2022}, though with significantly more involved implementations.

Quantum algorithms for solving PDEs that do not depend on a QLSA have also been developed. \citeauthor{Costa2019} \cite{Costa2019} put forward a quantum algorithm that evolves a quantum state according to the wave equation using Hamiltonian simulation without the need for a QLSA, aside from in the proposed generation of the initial conditions. A practical implementation of this algorithm was then developed and analysed by \citeauthor{Suau2021} \cite{Suau2021}, confirming that the gate requirements agreed with the theoretical complexity. In a different study, \citeauthor{Budinski2021} \cite{Budinski2021} proposed a quantum algorithm for the advection-diffusion equation centred around the lattice Boltzmann method (LBM) that tracks the evolution of particle distribution functions on a grid. The main challenge of implementing the non-unitary collision step of the LBM was achieved with the LCU method. Simulating ODEs in the form of Eq.\ \eqref{eq:ODE} as Hamiltonian simulations has received much recent attention \cite{Jin2022, Jin2023, Jin2024, An2023PhysRevLett, An2023}. \citeauthor{An2023PhysRevLett} \cite{An2023PhysRevLett} expressed the evolution as a linear combination of Hamiltonian simulation (LCHS) problems, which is a special case of the LCU method. The algorithm was later improved upon \cite{An2023} by further generalising the method and leading to the discovery of faster-decaying coefficients in the weighted sum, improving the $\epsilon$ dependency. \citeauthor{Jin2022} \cite{Jin2022, Jin2023, Jin2024} provided an alternative viewpoint on this problem by introducing a Schr\"{o}dingerisation method that maps linear PDEs to a higher-dimensional system of Schr\"{o}dinger equations, then solving with Hamiltonian simulation. Both the Schr\"{o}dingerisation \cite{Jin2022, Jin2023, Jin2024} and the LCHS \cite{An2023PhysRevLett, An2023} algorithms assume the Hermitian part of $M$ to be negative semi-definite, applicable to problems and numerical schemes that maintain or dampen the solution, but do not amplify it. This applies to upwind and central finite difference schemes for discretising the advection equation, but not downwind schemes. A time-marching algorithm for the ODE problem was proposed by \citeauthor{Fang2023} \cite{Fang2023} by explicitly integrating the PDE over short time steps, as is a common practice in the classical numerical solution of differential equations. This operator is not unitary so has a probability of failure in a block-encoding strategy, ordinarily leading to an exponentially decaying success probability in the simulation time $T$. This was overcome by applying a uniform singular value amplification at each time step, resulting in a runtime with a quadratic dependence on $T$. The absence of a QLSA in these methods generally leads to improved state preparation costs \cite{An2023} and favourable poly-logarithmic scaling in the allowable error per time step using sparse Hamiltonian simulation \citep{Berry2015} or LCU \cite{Childs2017} algorithms. Other methods of quantum matrix multiplication with the potential to be applied to solving PDEs using explicit time advancement were compared by \citeauthor{Shao2018} \cite{Shao2018} based on the swap test, singular value estimation and HHL algorithms. However, all of these methods utilise QPE as a subroutine, thus limiting their $\epsilon$ dependence and the number of state initialisation queries required.

The algorithm presented here uses an explicit time marching strategy for solving the advection equation by embedding numerical integrators into a series of Hamiltonian simulations. The algorithm achieves linear scaling in the required number of time steps $N_T$, requires a single copy of the initial quantum state and applies to various boundary conditions with arbitrary finite difference stencils. The mathematical description of the algorithm is provided next, followed by analyses of the errors and complexity in Sections \ref{sec:error} and \ref{sec:complexity} respectively. Statevector simulations of a two-dimensional laminar channel flow and a lid-driven cavity problem are provided and analysed in Section \ref{sec:simulations} and the paper is finalised with concluding remarks in Section \ref{sec:conclusions}.

\section{The Algorithm}
\label{sec:algorithm}

The first step is to discretise the advection equation in space and time using the FDM. Central, forward or backward schemes of any order of accuracy can be chosen for the spatial derivatives, and these can vary throughout the domain, e.g.\ by reducing the order of accuracy and transitioning to a one-sided scheme towards a wall. A second-order central scheme for a one-dimensional problem will be chosen to describe the algorithm and analyse its baseline properties, with different schemes being demonstrated later in the paper. When combined with forward Euler discretisation in time, the advection equation in one dimension for a constant velocity $u$ becomes
\begin{equation}
\frac{\phi_m^{t+1} - \phi_m^t }{\Delta t} + u\frac{\phi^t_{m+1} - \phi^t_{m-1}}{2\Delta x} = 0
\label{eq:discretised_adv}
\end{equation}
for spatial grid point $m$ and temporal location $t$. When solved classically, the forward-time central-space scheme in Eq.\ \eqref{eq:discretised_adv} is unstable for hyperbolic PDEs such as the advection equation, requiring upwind schemes for conditional stability \cite{Mitchell1980}. The quantum representation of such forward-time schemes in the present algorithm is stable for arbitrary finite difference stencils, as will be demonstrated later in this section. Equation \eqref{eq:discretised_adv} can be solved for $\phi_m^{t+1}$, obtaining an equation to advance the solution in time. A vector $\vec{\phi}_t = [\phi_0^t, \phi_1^t, \dots, \phi_{N-1}^t]$ can be constructed from $\phi$ to write Eq.\ \eqref{eq:discretised_adv} as a matrix transformation
\begin{equation}
    \vec{\phi}_{t+1} = A\vec{\phi}_t 
    \label{eq:matrix_system}
\end{equation}
For the described one-dimensional advection equation discretised with a second-order central FDM, the matrix $A$ takes the form
\begin{equation}
    A = 
    \begin{bmatrix}
        1   & -\frac{r}{2}       &    &  0    & \frac{r}{2}   \\
        \frac{r}{2}  & \ddots    & \ddots    & &  0  \\
         & \ddots    & \ddots    & \ddots &  \\
        0 & & \ddots & \ddots & -\frac{r}{2} \\
        -\frac{r}{2}   &     0   &    & \frac{r}{2}        & 1
      \end{bmatrix}
    \label{eq:A_matrix}
\end{equation}
when considering simple periodic boundary treatment where the first and last grid points are adjacent, leading to entries in the top-right and bottom-left corners. The stability parameter $r=u\Delta t/\Delta x$ is the Courant-Friedrichs-Lewy (CFL) number and is related to the condition number $\kappa$. To ensure numerical stability, $r$ must not exceed $1$ \citep{Courant1928}, though in practice and especially for explicit schemes, much lower values are required. In $D$-dimensional space with $k$-order-accurate spatial discretisation, the sparsity of the matrix $s=1+Dk$ with FDM coefficients determined by $k$. 

\begin{figure}[t]
    \centering
    \includegraphics[width=0.65\columnwidth]{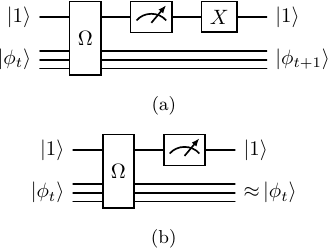}
    \caption{Quantum circuit for (a) a successful time step measuring $\ket{0}$ and (b) an unsuccessful time step measuring $\ket{1}$.}
    \label{fig:quantum_circuit}
\end{figure}

While divergence-free advection is a norm-preserving process, the truncation errors from the discretisation procedure in Eq.\ \eqref{eq:discretised_adv} lead to the matrix $A$ being non-unitary, i.e.\ $A^\dagger A \ne I$ and $AA^\dagger \ne I$. Therefore, to enact $A$ on a quantum state $\ket{\phi_t}$, the non-unitary Hamiltonian embedding procedure described by \citeauthor{Gingrich2004} \cite{Gingrich2004} is used. A Hamiltonian $H$ is constructed from $A$ in the form
\begin{equation}
    H =
    \begin{bmatrix}
        0 & iA \\
        -iA^\dagger & 0
    \end{bmatrix}
    \label{eq:hamiltonian_embedding}
\end{equation}
where $i$ is the imaginary unit. A quantum state is evolved according to $H$ by the unitary operator
\begin{align}
    \Omega &= \exp\left(-iH\theta \right) \nonumber \\
    &= \exp
    \begin{bmatrix}
        0 & A\theta \\
        -A^{\dagger}\theta & 0
    \end{bmatrix}
    \label{eq:omega_definition}
\end{align}
where the symbol $\theta$ is the Hamiltonian evolution time for a single time step. The time step size is encoded in the Hamiltonian, so $\theta$ affects the accuracy of a time step and its probability of success rather than the evolution time itself. The exponential function produces a block matrix with the structure \cite{Gingrich2004}
\begin{align}
    \exp\begin{bmatrix} 
        0 & X \\ Y & 0 
    \end{bmatrix} &= 
    \begin{bmatrix} \cosh(\sqrt{XY}) & X\frac{\sinh(\sqrt{YX})}{\sqrt{YX}} \\ Y\frac{\sinh(\sqrt{XY})}{\sqrt{XY}} & \cosh(\sqrt{YX}) 
    \end{bmatrix} \nonumber \\
    &= 
    \begin{bmatrix} 
        \cos(\theta\sqrt{AA^\dagger}) & A\frac{\sin(\theta\sqrt{A^\dagger A})}{\sqrt{A^\dagger A}} \\ -A^\dagger \frac{\sin(\theta\sqrt{AA^\dagger})}{\sqrt{AA^\dagger}} & \cos(\theta\sqrt{A^\dagger A}) 
    \end{bmatrix} \label{eq:sinh_cosh}
\end{align}
where the top-right and top-left blocks of Eq.\ \eqref{eq:sinh_cosh} are termed $\widetilde{A}$ and $\widetilde{I}$ respectively. The $\widetilde{A}$ matrix is closely proportional to $A$ for small values of $\theta$ or when $\sqrt{A^\dagger A} \approx I$ since $\sin(I\theta) = I\sin(\theta)$, leading to $ \widetilde{A} \approx A\sin(\theta)$. Furthermore, the $\widetilde{I}$ matrix is closely proportional to $I$ for small values of $\theta$ or when $\sqrt{AA^\dagger} \approx I$, leading to $\widetilde{I} \approx I\cos(\theta)$. For the advection equation evolution operator in Eq.\ \eqref{eq:A_matrix}, $\sqrt{A^\dagger A}$ and $\sqrt{AA^\dagger}$ can be approximated by $I + O(r^2)$ for small values of $r$. Equation \eqref{eq:sinh_cosh} also reveals that the singular values $\sigma_i(\widetilde{A}) = \sin(\sigma_i(A)\theta)$ are bounded by 1, stabilising the scheme regardless of the finite difference stencil. Applying $\Omega$ to a solution register $\ket{\phi_t}$ supplemented by an ancilla qubit initialised as $\ket{1}$ produces the state
\begin{align}
    \Omega \ket{1}\ket{\phi_t} = \Omega 
     \begin{bmatrix}
        0 \\
        \ket{\phi_{t}}
    \end{bmatrix}
    &= 
    \begin{bmatrix}
        \widetilde{A} \ket{\phi_{t}} \\
        \widetilde{I} \ket{\phi_{t}}
    \end{bmatrix} \label{eq:state_evolution} \\
    &\approx
    \begin{bmatrix}
        \ket{\phi_{t+1}} \\
        \ket{\phi_{t}}
    \end{bmatrix} \nonumber
\end{align}
Postselecting the ancilla qubit in the state $\ket{0}$ collapses the solution register to $\ket{\phi_{t+1}} = \widetilde{A}\ket{\phi_t}$, and if postselection fails, the state collapses to $\widetilde{I} \ket{\phi_t} \approx \ket{\phi_t}$, allowing the state to be reused for further computation. The quantum circuit representations of a successful and unsuccessful time step are shown in Fig.\ \ref{fig:quantum_circuit}. 

\section{Error Analysis}
\label{sec:error}

An expression for the error upper bound can be derived by considering the errors at each time step associated with the application of $\widetilde{A}$ and $\widetilde{I}$, and combining them with the number of time steps required and the probability of postselection success. First, the errors per time step are derived in Part A, followed by the probability of a successful time step in Part B. These are combined to produce the overall error bound in Part C and compared to the errors when applied to a different PDE, the heat equation, in Part D.

\subsection{Error per Time Step}
\label{sec:per_timestep}

The matrix $A$ in Eq.\ \eqref{eq:A_matrix} is a Toeplitz matrix so can be defined in terms of its diagonals, where the main diagonal $d_0 = 1$, super-diagonal $d_{1} = -r/2$ and sub-diagonal $d_{-1}=r/2$. Similarly, the corresponding terms in $\widetilde{A}$ can be evaluated as a function of $r$ and $\theta$. Considering a $4\times 4$ matrix to avoid unnecessary negligible terms, $\widetilde{A}$ can be expressed as
\begin{align}
    \widetilde{A} &= \nonumber \\
    d_0&: \frac{1}{2} \left(\sin (\theta )+\frac{\sin \left(\theta  \sqrt{r^2+1}\right)}{\sqrt{r^2+1}}\right) \nonumber\\
    d_{1},-d_{-1},&: -\frac{r \sin \left(\theta  \sqrt{r^2+1}\right)}{2 \sqrt{r^2+1}} \label{eq:AThetaTilde} \\
    d_2, d_{-2}&: \frac{1}{2} \left(\sin (\theta )-\frac{\sin \left(\theta  \sqrt{r^2+1}\right)}{\sqrt{r^2+1}}\right) \nonumber
\end{align}
The similarity between the matrices in Eq.\ \eqref{eq:A_matrix} and \eqref{eq:AThetaTilde} can be compared by constructing an error matrix that quantifies the error of the terms relative to the main diagonal. The property that Eq.\ \eqref{eq:AThetaTilde} must satisfy is $d_1, -d_{-1} = -rd_0/2$ with other elements equalling 0, i.e.\ the relative proportions of the matrix must be consistent. If Eq.\ \eqref{eq:A_matrix} is scaled by $d_0^{(\text{Eq}.\ref{eq:AThetaTilde})}$ to yield equal diagonal elements with Eq.\ \eqref{eq:AThetaTilde}, then subtracting Eq.\ \eqref{eq:AThetaTilde} from the scaled Eq.\ \eqref{eq:A_matrix} gives the error matrix $E_A$, given by
\begin{align}
    E_A &= \nonumber \\
    d_0&: 0 \label{eq:sin_error_matrix}\\
    d_{1}, -d_{-1}&: -\frac{r}{4} \left(\sin (\theta )-\frac{\sin \left(\theta  \sqrt{r^2+1}\right)}{\sqrt{r^2+1}}\right) \nonumber \\
    d_2, d_{-2}&: -\frac{1}{2} \left(\sin (\theta )-\frac{\sin \left(\theta  \sqrt{r^2+1}\right)}{\sqrt{r^2+1}}\right) \nonumber
\end{align}
The maximum error for the application of $\widetilde{A}$ can then be defined by taking the spectral norm of Eq.\ \eqref{eq:sin_error_matrix}
\begin{equation}
    || E_A || = \frac{1}{2} \left(\sin (\theta )\sqrt{r^2+1} -\sin \big(\theta  \sqrt{r^2+1}\big)\right)
    \label{eq:EA}
\end{equation}
Following the same procedure, $\widetilde{I}$ can be written as
\begin{align}
    \widetilde{I} &= \nonumber \\
    d_0&: \frac{1}{2} \left(\cos (\theta )+\cos \left(\theta  \sqrt{r^2+1}\right)\right) \nonumber\\
    d_{-1},-d_1,&: 0 \label{eq:ITilde}\\
    d_2, d_{-2}&: \frac{1}{2} \left(\cos (\theta )-\cos \left(\theta  \sqrt{r^2+1}\right)\right) \nonumber
\end{align}
and an error matrix can be constructed that quantifies error relative to the identity matrix:
\begin{align}
    E_I &= \nonumber \\
    d_0&: 0 \label{eq:cos_error_matrix}\\
    d_{1}, d_{-1}&: 0 \nonumber \\
    d_2, d_{-2}&: -\frac{1}{2} \left(\cos (\theta )-\cos \left(\theta  \sqrt{r^2+1}\right)\right)  \nonumber
\end{align}
Taking the spectral norm of the error matrix gives
\begin{equation}
    ||E_I|| = \frac{1}{2} \left( \cos (\theta )-\cos \left(\theta\sqrt{r^2+1} \right)\right)
    \label{eq:EI}
\end{equation}
which represents the maximum error for a failed postselection. 

\subsection{Time Step Success Probability}
\label{sec:probability}

\begin{figure}[bt!]
    \centering
    \includegraphics[width=\columnwidth]{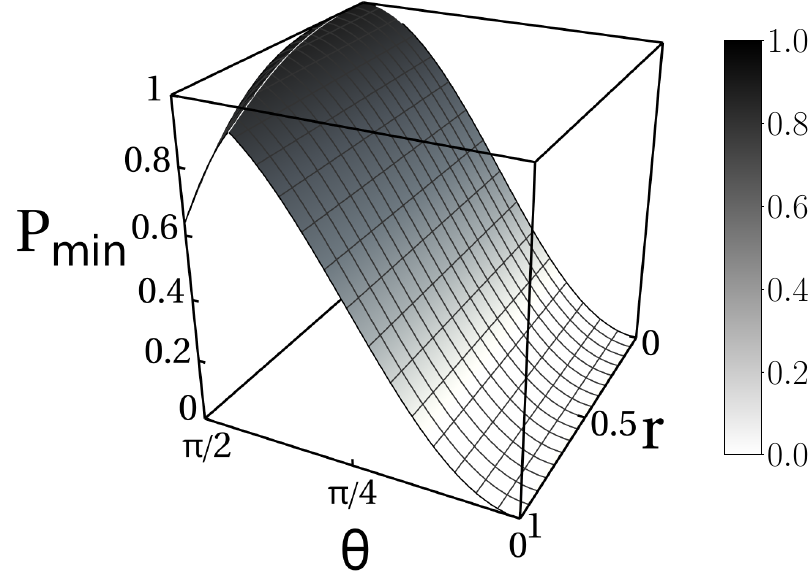}
    \caption{Surface plot of the minimum probability of a successful time step defined in Eq.\ \eqref{eq:Pmin}.}
    \label{fig:Pmin}
\end{figure}

The probability of a successful time step $P=||\widetilde{A}\ket{\phi}||^2$ can be studied by analysing the contribution of $\widetilde{I}$ to the state. The square of the spectral norm $||\widetilde{I}||^2$ corresponds to the largest action of $\widetilde{I}$ on a statevector squared, so can be used to find the worst-case probability of successful measurement $P_\text{min} = 1 - ||\widetilde{I}||^2$. Using the definition of $\widetilde{I}$ in Eq.\ \eqref{eq:ITilde}, $P_\text{min}$ is defined as
\begin{equation}
    P_\text{min} = 
    \begin{cases}
        \sin^2(\theta) & \text{ when } 0 < \theta \leq \frac{\pi}{1+\sqrt{r^2+1}} \\
        \sin^2(\theta \sqrt{r^2+1}) & \text{ when } \frac{\pi}{1+\sqrt{r^2+1}} < \theta \leq \frac{\pi}{2}
    \end{cases}
    \label{eq:Pmin}
\end{equation}
which is visualised in Fig.\ \ref{fig:Pmin}. The figure shows that $P_\text{min}$ is optimal for small values of $r$ and when $\theta=\pi/(1+\sqrt{r^2+1})$, approaching unity as $r$ approaches zero. Using this value of $\theta$ for a typical case where $r=0.1$, the minimum probability $P_\text{min} = 99.9985 \%$, corresponding to a worst case of $67\,000$ successful time steps per failed time step on average. It is shown in Section \ref{sec:simulations} that for a practical configuration, the probability of postselection success is mostly represented by $\sin^2(\theta)$, with the contribution from $\sin^2(\theta\sqrt{r^2+1})$ having a reduced role leading to $\theta=\pi/2$ being optimal in practice. For example, $r=0.1$ and $\theta=\pi/2$ resulted in approximately 1 billion successful time steps for every failed time step.

\subsection{Overall Error Bound}
\label{sec:overall_error}

\begin{figure*}[bt!]
    \centering
    \includegraphics[width=0.75\textwidth]{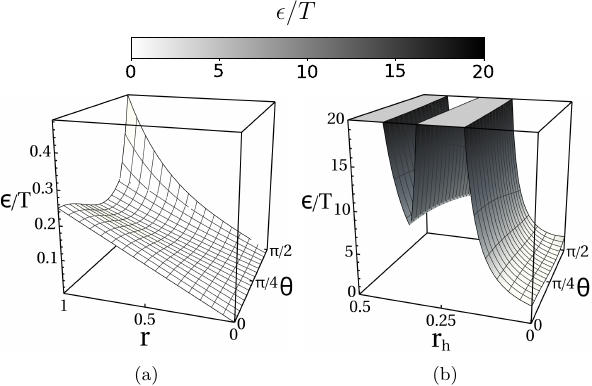}
    \caption{Surface plot of the error bound per non-dimensional simulation time as a function of $r$ and $\theta$ for (a) the advection equation and (b) the heat equation. For the heat equation in (b), the surface has been truncated at $\epsilon/T = 20$ since $\epsilon$ asymptotically approaches infinity for $r_h\to 0.25$ and $r_h\to 0.5$.}
    \label{fig:epsilon_plot}
\end{figure*}

The overall error bound for the algorithm can be evaluated using $\epsilon = N_T||E_A|| + N_F||E_I||$, where $N_T$ is the required number of successful time steps and $N_F$ is the expected number of failed time steps. Since the number of time steps required is inversely proportional to the time step, $N_T=T/r$, where $T$ is the simulation time in non-dimensional form and $N_F=N_T/P_\text{min}-N_T$. Combining these expressions, the error per the simulation time evaluates to
\begin{align}
    \frac{\epsilon}{T} = 
    \begin{cases}
    \dfrac{1}{2r}
    \begin{bmatrix}
        \left(\cos (\theta )-\cos \left(\theta  \sqrt{r^2+1}\right)\right)\cot ^2(\theta) \\
        +\sin (\theta )\sqrt{r^2+1} -\sin \left(\theta  \sqrt{r^2+1}\right)
    \end{bmatrix} \\[4mm]
    \dfrac{1}{2r}
    \begin{bmatrix}
        \left(\cos (\theta )-\cos \left(\theta  \sqrt{r^2+1}\right)\right) \cot ^2\left(\theta  \sqrt{r^2+1}\right) \\
        +\sin (\theta )\sqrt{r^2+1} -\sin \left(\theta  \sqrt{r^2+1}\right)
    \end{bmatrix}
    \end{cases}
    \label{eq:error_bound}
\end{align}
when $0 < \theta \leq \frac{\pi}{1+\sqrt{r^2+1}}$ and $\frac{\pi}{1+\sqrt{r^2+1}} < \theta \leq \frac{\pi}{2}$ respectively. This bound assumes that the number of failed time steps is relatively close to the expected value, which is reasonable as $N_T$ becomes large. This expression is visualised in Fig.\ \ref{fig:epsilon_plot}a, which reveals that the overall error is mostly insensitive to the value of $\theta$ and grows linearly with $r$. This indicates that for the low values of $r$ required for numerical stability, $\theta=\pi/2$ is the most efficient configuration as it minimises the required circuit depth without incurring substantial additional errors. The linear growth of error in the quantum matrix representation indicates that the algorithm does not worsen the error complexity from the classical Euler method that underpins the algorithm, and means there is little algorithmic benefit in pursuing higher-order time integrators.

\subsection{Comparison with the Heat Equation}
\label{sec:heat_equation_comparison}

To demonstrate the advantageous properties of Eq.\ \eqref{eq:error_bound} for the advection equation, the error will be compared against the algorithm applied to a different PDE, the heat equation. The heat equation features a second derivative term on the right-hand side
\begin{equation}
    \frac{\partial \phi}{\partial t} = D\frac{\partial^2\phi}{\partial x_j\partial x_j}
\end{equation}
where $D$ is the diffusivity. Following the same discretisation procedure for a one-dimensional problem, the matrix $A$ takes the form $[d_{-1}, d_0, d_1] = [r_h, 1-2r_h, r_h]$ for the internal grid points, where $r_h = D\Delta t / (\Delta x)^2$ is the stability parameter with a theoretical maximum value of 0.5. The error bound for the heat equation evaluates to
\begin{align}
    \dfrac{\epsilon}{T} {=} 
    \begin{cases}
        \dfrac{1}{2r_h}
        \begin{bmatrix*}[l]
            \dfrac{\begin{vmatrix*}[l](8r_h{-}3)\sin(\theta)+\sin(\theta{-}4r_h\theta) \\ +2\sin(\theta{-}2r_h\theta)\end{vmatrix*}}{2{-}4r_h} \\
            +\left|\sin (\theta {-}3r_h\theta) + 3\sin(\theta {-} r_h\theta) \right| \\
            \times\cot ^2(\theta {-}4r_h \theta)\sin(r_h\theta )
        \end{bmatrix*} \\[8mm]
        \dfrac{1}{2r_h}
        \begin{bmatrix*}[l]
            \dfrac{\begin{vmatrix*}[l](1{-}4r_h)\sin(\theta) +(4r_h{-}3)\sin(\theta{-}4r_h\theta) \\ +(2{-}8r_h)\sin(\theta{-}2r_h\theta)\end{vmatrix*}}{2{-}4r_h} \\
            +\left|\sin (\theta {-}3r_h\theta) + 3\sin(\theta {-} r_h\theta) \right| \\
            \times\cot ^2(\theta {-}2r_h \theta)\sin(r_h\theta )
        \end{bmatrix*}
    \end{cases}
\end{align}
when $0<r\leq 1/3$ and $1/3<r\leq 1/2$ respectively, which is visualised in Fig.\ \ref{fig:epsilon_plot}b. The expression and the figure reveal that the error asymptotically approaches infinity for $r_h\to 0.25$ and $r_h\to 0.5$. Furthermore, as $r_h\to 0$, the error bound does not approach zero as occurs for the advection equation but rather approaches approximately 2, thereby making it impossible to diminish the error by increasing the circuit depth. A local minimum value for the error occurs when $r_h=1/3$, where $\epsilon/T$ reduces to approximately 6 for $\theta = \pi/2$. The errors for the advection equation are analogous to classical computations since reducing $r$ linearly reduces the error, with the error approaching zero as $r\to 0$. This is not the case when applied to the heat equation where the error is a convoluted, discontinuous function of $r_h$ without a clear trend. Even using the optimal values of $r_h=1/3$ or $r_h < 0.01$, the errors are still too large for useful computation, indicating that the heat equation evolution operator cannot be efficiently represented in this manner. Therefore, the algorithm cannot be considered a sufficiently general PDE solver.

\section{Complexity Analysis}
\label{sec:complexity}

The number of qubits grows as $n = O(\log N)$ since the computational grid is compressed into the amplitudes of the exponentially growing computational basis states. The circuit depth grows linearly with the required number of time steps $N_T \propto TN^{1/D}$. The number of time steps growing linearly with the desired simulation time $T$ is intuitive, but the $N^{1/D}$ factor with the number of grid points $N$ requires a further breakdown. It is assumed that a $D$-dimensional domain of equal side lengths is discretised by $N = N_x^D$ grid points, where $N_x=N_y=N_z$ for $D>1$. The CFL number $r_\text{max} = \text{max}(u_{j})\Delta t/\Delta x$ constrains the time step size to the grid spacing, where halving $\Delta x$ requires halving $\Delta t$, or equivalently, doubling $N_x$ requires doubling $N_T$. Given that $N_T\propto N_x$ and $N = N_x^D$, then $N \propto N_T^D$, or inversely $N_T \propto N^{1/D}$.

In terms of error, Fig.\ \ref{fig:epsilon_plot} shows that $\epsilon$ grows linearly with $r$, which is the same as the underpinning Euler method. Halving $r$ halves $\epsilon$, which requires doubling $N_T$ and therefore doubling the circuit depth, so the circuit depth grows with $1/\epsilon$.  

\begin{figure*}
    \centering
    \includegraphics[width=0.86\textwidth]{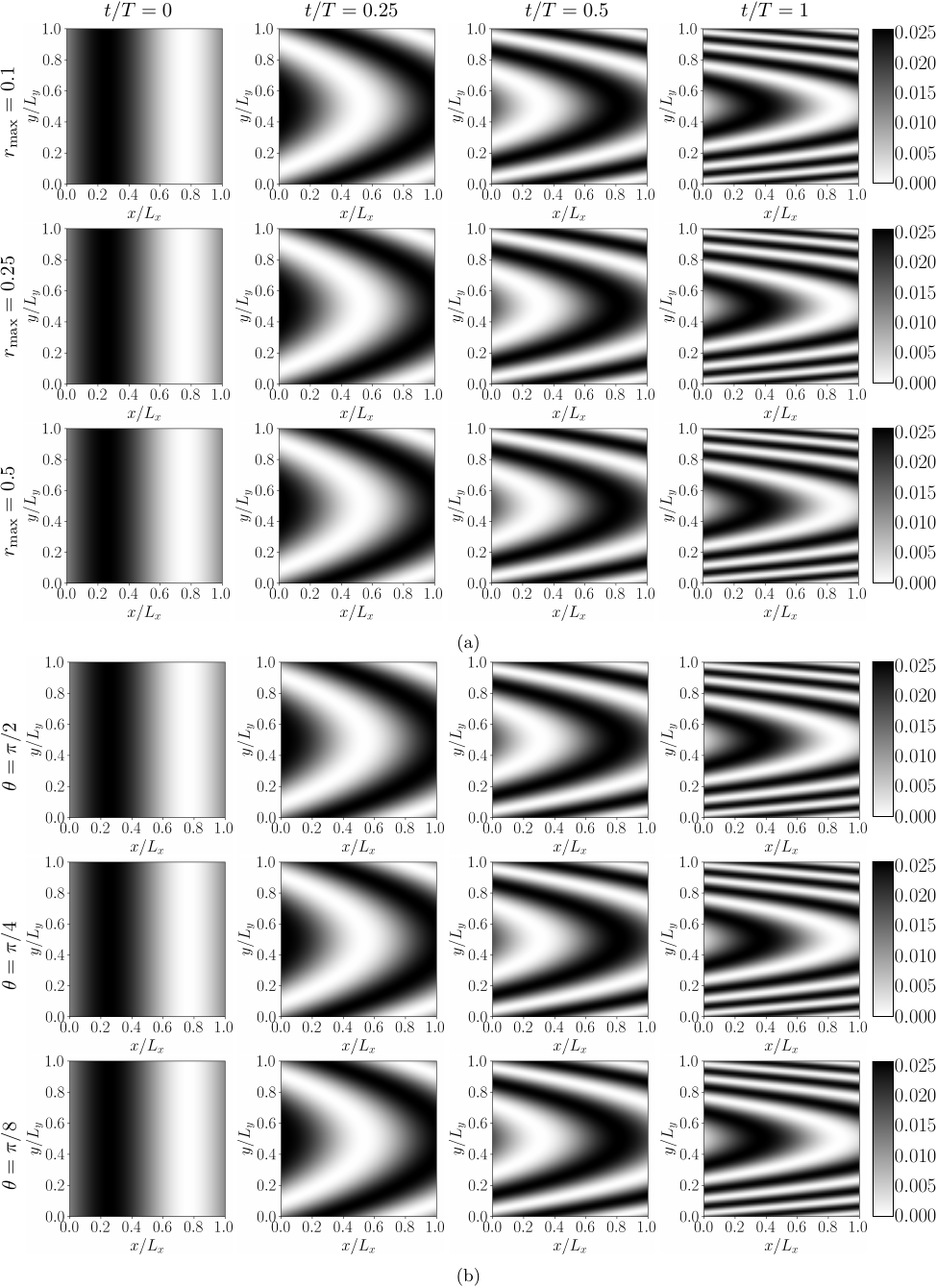}
    \caption{Evolution of the quantum amplitudes from statevector simulations of a 2D laminar channel flow described by the advection equation at different time intervals for (a) $\theta = \pi/(1+\sqrt{r_\text{max}^2+1})$ while varying $r_\text{max}$ and (b) $r_\text{max}=0.25$ while varying $\theta$. The simulations use a 2\textsuperscript{nd}-order-accurate central finite difference stencil.}
    \label{fig:advection_evolution}
\end{figure*}

\begin{figure*}[htb]
  \centering
  \includegraphics[width=0.8\linewidth]{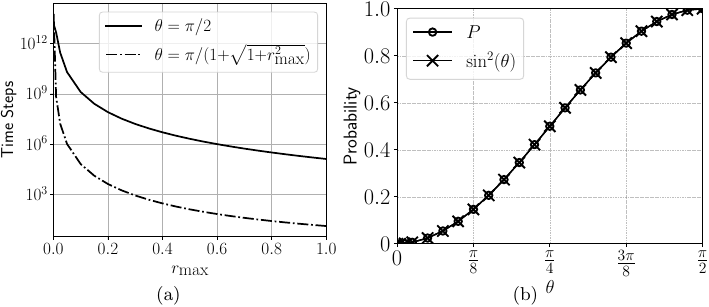}
  \caption{(a) The typical number of successful time steps for every failed time step as a function of $r_\text{max}$ for different values of $\theta$, and (b) the probability of a successful time step $P=|| \widetilde{A}\ket{\phi_t} ||^2$ and the prediction $\sin^2(\theta)$ for the $r_\text{max}=0.25$ case.}
  \label{fig:probability_omega}
\end{figure*}

The remaining complexity considerations depend on the chosen implementation of the Hamiltonian simulations. As an example, the algorithm of \citeauthor{Berry2015} \cite{Berry2015} is considered due to its near-optimal properties. It combines the strategies of a Szegedy quantum walk \cite{Childs2010, Berry2009} and fractional-query simulation \cite{Berry2014exponential}. The Szegedy quantum walk approach scales optimally in the matrix sparsity but not in the allowable error, while the simulation of the fractional-query model scales optimally with the error but not with the sparsity. The fractional query model is used to correct the phase more accurately than the QPE step in the quantum walk approaches \cite{Childs2010, Berry2009}, resulting in the favourable scaling in both sparsity and the allowable error. The overall algorithm implements Hamiltonian simulation with
\begin{equation}
    O\left( \tau\left(n + \log^{5/2}(\tau/\epsilon) \right) \frac{\log(\tau/\epsilon)}{\log \log (\tau/\epsilon)} \right)
    \label{eq:hamiltonian_efficiency}
\end{equation}
gates, where $\tau = s ||H||_\text{max}\theta$, $s$ is the sparsity and $||H||_\text{max}\theta$ is the maximum value of the matrix $H\theta$ \cite{Berry2015} which is $O(1)$ in this case. The complexity of the Hamiltonian simulation step can therefore be written as
\begin{equation}
    O\left( s \left(\log(N) + \log^{5/2}(s/\epsilon) \right) \frac{\log(s/\epsilon)}{\log \log (s/\epsilon)} \right)
\end{equation}
Suppressing the poly-logarithmic terms for simplicity, this reduces to an almost linear dependence on the sparsity, $\widetilde{O}(s)$. For the block-encoded Hamiltonian in Eq.\ \eqref{eq:hamiltonian_embedding}, the sparsity of $H$ is equal to the sparsity of $A$, which is $s = 1 + Dk$ where $k$ is the order of the spatial discretisation. Therefore, the complexity of the Hamiltonian simulation step can be written as $\widetilde{O}(Dk)$. Since the error $\epsilon$ does not appear in this simplified expression, the dominant source of error arises from the encoding of the non-unitary operator and the explicit Euler method rather than the Hamiltonian simulation implementation. Combining all of these contributions, the circuit depth grows as $\widetilde{O}(N_Ts/\epsilon)$ or $\widetilde{O}\left(TN^{1/D}Dk/\epsilon \right)$ for $D$-dimensional simulations.

Efficient classical simulations of the advection equation typically have a time complexity of $O(NN_T)$ \cite{Pope2000}, and given that $N_T \propto TN^{1/D}$, this can be written as $O(TN^{(1+D)/D})$. The quantum algorithm offers a significant polynomial improvement over the classical algorithm as their time complexities differ by a factor of $N$. The effectiveness of the quantum algorithm increases in higher dimensional space as the number of grid points increases at a much faster rate than the required circuit depth, taking advantage of the exponentially growing Hilbert space.

\section{Simulations}
\label{sec:simulations}

\begin{figure*}
    \centering
    \includegraphics[width=0.86\textwidth]{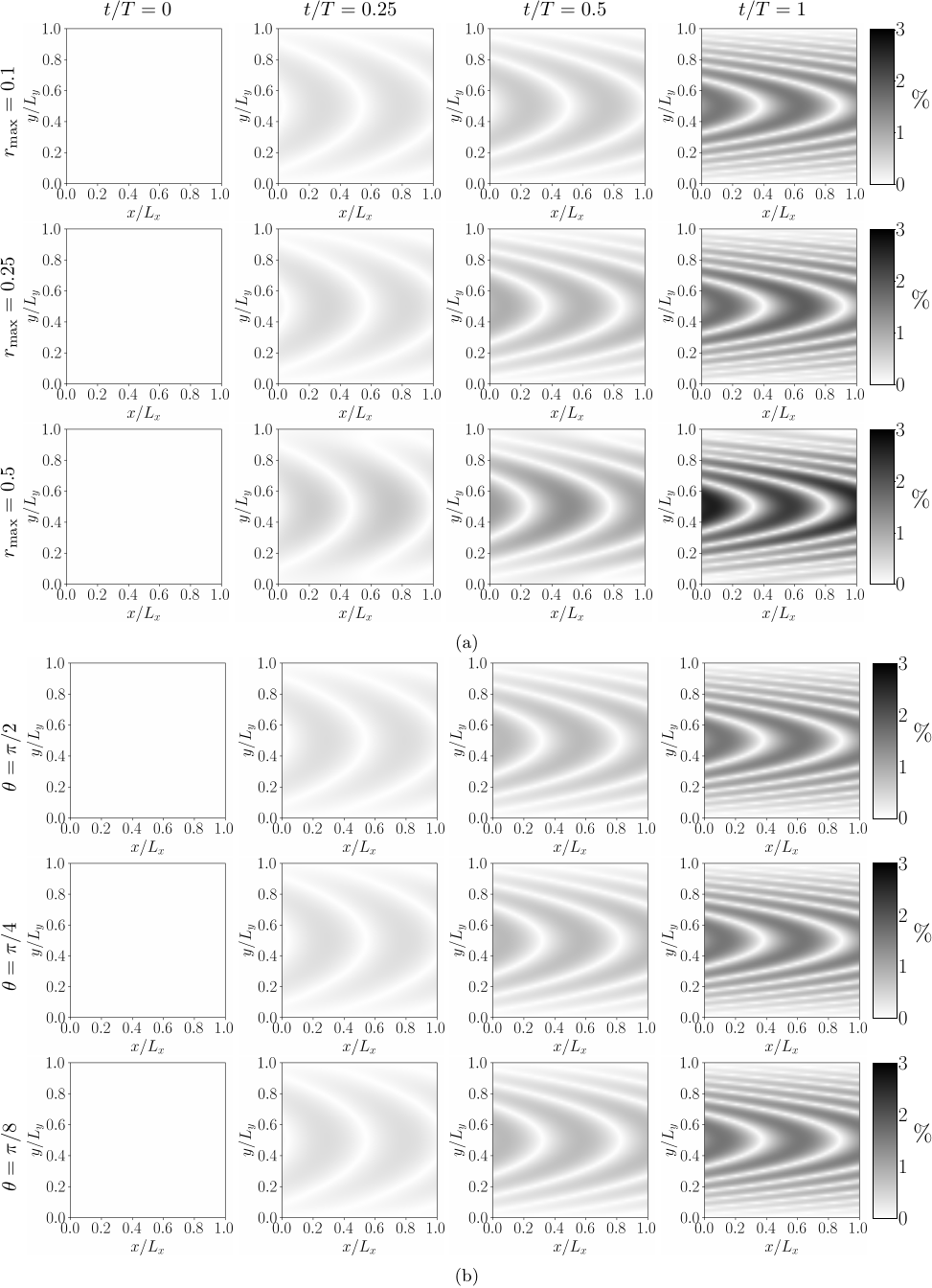}
    \caption{Error contours for the channel flow simulations compared to the analytical solution as a percentage, defined as $100\big|\phi(x,y,t)-\ket{\phi}\big|/\text{max}(\phi(x,y,t))$, for (a) $\theta = \pi/(1+\sqrt{r_\text{max}^2+1})$ while varying $r_\text{max}$ and (b) $r_\text{max}=0.25$ while varying $\theta$.}
    \label{fig:error_map}
\end{figure*}

This section presents quantum statevector simulations of the algorithm applied to a two-dimensional laminar channel flow in Part A and a lid-driven cavity flow in Part B. In Part C, the response of the algorithm to noise in the initial quantum state and the Hamiltonian embedding is demonstrated, and the errors are quantified for various finite difference stencils with and without noise. In all simulations, $N=64\times 64=4096$ grid points have been used to discretise the problems in space, corresponding to 12 qubits representing the solution and the additional ancilla qubit required by the algorithm, totalling 13 qubits.

\subsection{Laminar Channel Flow}

The velocity field $\vec{u}=[u,v]$ is described by the analytical solution to the Navier-Stokes equations in this configuration, known as a plane Poiseuille flow, which has a parabolic profile of $u$ leading to the non-dimensional CFL parameter to be defined as
\begin{align}
    r(y) &= \frac{u(y)\Delta t}{\Delta x} \\
    &= r_\text{max}\left(4y(1-y)\right) \quad 0\leq y \leq 1
\end{align}
which varies from from 0 at the walls to $r_\text{max}$ at the centre of the domain, where $y$ is the vertical normalised distance. The vertical component of the velocity $v=0$ because the flow is laminar.  The scalar $\phi$ is initialised as a sine wave in the horizontal direction as
\begin{equation}
    \phi(x)=\sin(2\pi x)+1, \quad 0\leq x \leq 1
    \label{eq:initial_conditions}
\end{equation}
where $x$ is the corresponding normalised horizontal distance. The $x$ boundaries are periodic, so fluid that flows out of the right boundary enters through the left boundary. The velocity reducing to zero at the walls (i.e.\ a no-slip wall) corresponds to Dirichlet boundary conditions, where the value of $\phi_\text{wall}$ is maintained according to Eq.\ \eqref{eq:initial_conditions}. This boundary condition results in a single entry of $1$ on the diagonal of $A$.

\begin{figure*}
    \centering
    \includegraphics[width=\textwidth]{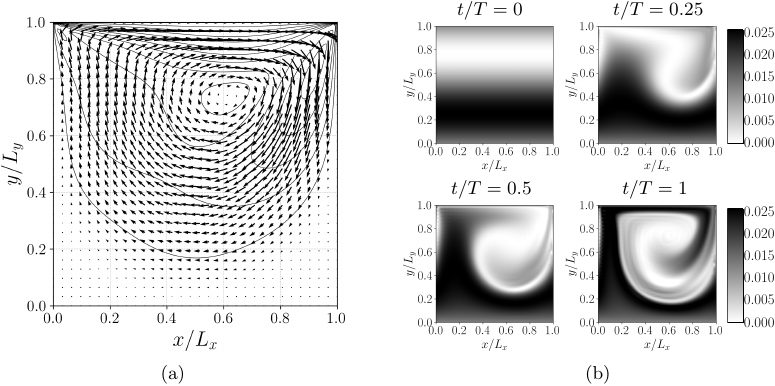}
    \caption{(a) Velocity vectors overlaid with velocity magnitude contours for the lid-driven cavity simulations at $Re=100$, and (b) contours showing the evolution of the quantum amplitudes for lid-driven cavity simulations discretised by a one-sided 2\textsuperscript{nd}-order-accurate upwind finite difference stencil.}
    \label{fig:lid_driven_cavity}
\end{figure*}
\begin{figure*}
    \centering
    \includegraphics[width=\textwidth]{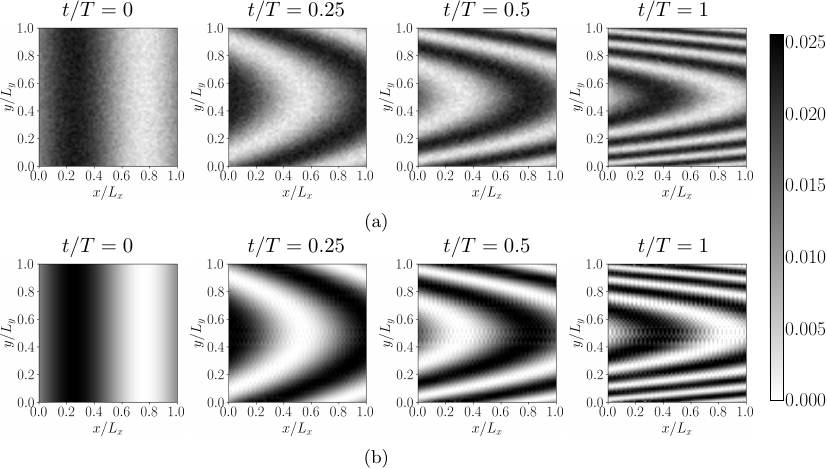}
    \caption{Evolution of the quantum amplitudes from statevector simulations of a noisy channel flow with (a) Gaussian noise in the initial state and (b) Gaussian noise in the Hamiltonian embedding. The simulations use a 4\textsuperscript{th}-order-accurate central finite difference stencil.}
    \label{fig:noise_contours}
\end{figure*}

The evolution of the quantum amplitudes for (a) varying CFL parameter $r$ and (b) varying Hamiltonian evolution time $\theta$ per time step is provided in Fig.\ \ref{fig:advection_evolution}, showing that a visually similar solution is reached regardless of the values of $r$ or $\theta$. The equivalent simulation time corresponds to 2000 successful time steps for $r_\text{max}=0.1$, 800 successful time steps for $r_\text{max}=0.25$ and 400 successful time steps for $r_\text{max}=0.5$. The simulations in Fig.\ \ref{fig:advection_evolution}a use $\theta = \pi/(1+\sqrt{r_\text{max}^2+1})$ calculated from Fig.\ \ref{fig:Pmin} to optimise $P_\text{min}$, i.e.\ the worst-case probability of measurement success. The simulations in \ref{fig:advection_evolution}b use a constant $r_\text{max}=0.25$ while varying $\theta$ to $\pi/2$, $\pi/4$ and $\pi/8$. Varying $\theta$ alters the probability of a successful time step $P=|| \widetilde{A}\ket{\phi_t} ||^2$, and the variation of $P$ with $r_\text{max}$ and $\theta$ is presented in Fig.\ \ref{fig:probability_omega}a and \ref{fig:probability_omega}b respectively. The value of $\theta$ that optimises the worst-case $P_\text{min}$ is not found to be optimal in practice, with $\pi/2$ providing a significantly greater probability of postselection success. For example, $r_\text{max}=0.1$ and $\theta=\pi/(1+\sqrt{r_\text{max}^2+1})$ leads to a typical 67\,000 successful time steps for every unsuccessful time step compared to approximately 1 billion when $\theta=\pi/2$. The prediction of $\sin^2(\theta)$ from Fig.\ \ref{fig:Pmin} is very accurate for all values of $\theta$ as shown in Fig.\ \ref{fig:probability_omega}b. Although the probability of time step success is close to certain when $\theta=\pi/2$, there remains a small chance of failure so the algorithm cannot be considered entirely deterministic. However this does not impact the ability of the algorithm to prepare the state $\ket{\phi_T}$, it just requires further attempted time steps. The $\theta=\pi/4$ and $\pi/8$ cases in Fig.\ \ref{fig:advection_evolution}b demonstrate the ability of the algorithm to withstand postselection failure, where approximately only 50\% and 14.6\% of the time steps succeed, respectively.

The laminar channel flow configuration can be considered as an ensemble of one-dimensional advection problems, so has an analytical solution that can be calculated using the method of characteristics. Given the initial condition in Eq.\ \eqref{eq:initial_conditions}, the analytical solution at time $t$ is
\begin{equation}
    \phi(x,y,t) = \frac{\sin(2\pi(x-u(y)t)) +1}{||\phi(x,y,0)||}
    \label{eq:analytical_solution}
\end{equation}
where the denominator ensures a norm of 1 so the solutions can be compared like-for-like. Figure \ref{fig:error_map} compares the quantum solution for all cases against the analytical solution by plotting the local errors as a percentage, $100\big|\phi(x,y,t)-\ket{\phi}\big|/\text{max}(\phi(x,y,t))$. The errors remain within 3\% for every case and appear to grow linearly with the local value of $r$, increasing towards the centre of the domain where $r$ is greatest and with increasing $r_\text{max}$ as shown in Fig.\ \ref{fig:error_map}a. This is in agreement with the theoretical linear dependency of the algorithm with the error derived in Section \ref{sec:complexity}. Figure \ref{fig:error_map}b shows how the errors vary with $\theta$ under the influence of unsuccessful time steps, which shows that reducing $\theta$ does not decrease the overall accuracy of the solution, again agreeing with the theoretical expression visualised in Fig.\ \ref{fig:epsilon_plot}. However, reducing $\theta$ does require more time steps to be attempted, thereby increasing the circuit depth and confirming the theoretical result that $\theta=\pi/2$ is the most efficient configuration. In all cases, the errors smoothly display the parabolic profiles of the scalar field, indicating that discretisation errors are consistently leading to propagation at a slightly different velocity to $u(y)$. The parabolas of zero error are false negatives, arising due to identical scalar values on either side of a local maximum or minimum masking the true error. The algorithm accurately represents the Dirichlet boundary where the scalar gradients are greatest, as demonstrated by the errors reducing to zero towards the $y$ boundaries.

\subsection{Lid-Driven Cavity}

A scalar transported in a lid-driven cavity has been simulated to demonstrate the performance of the algorithm for a multi-dimensional problem where no straightforward analytical solution exists. The lid-driven cavity configuration involves a square cavity filled with a fluid, where the top wall (lid) moves horizontally at a constant velocity $U_\text{wall}$, while the other three walls remain stationary. The fluid directly adjacent to the moving wall acquires the same velocity as the wall due to the no-slip condition. This imparts momentum to the rest of the fluid, causing it to circulate within the cavity.

As no analytical solution is available for this configuration and for comparison, the velocity field has been generated by the commercial computational fluid dynamics software Ansys Fluent \cite{Ansys2024} for a Reynolds number $\text{\itshape Re} = U_\text{wall}L/\nu = 100$, where $L$ is the side length of the cavity and $\nu$ is the kinematic viscosity. The domain was discretised with $64\times 64$ cells and solved using the finite volume method with a pressure-based coupled solver and a second-order upwind scheme for the spatial derivatives. The resulting velocity vectors are shown in Fig.\ \ref{fig:lid_driven_cavity}a, overlaid by the velocity magnitude. The vectors are drawn on a $32\times 32$ grid for a less-cluttered graphic.

The scalar field is initialised as a sine wave in the $y$ direction, $\phi(x,y) = \sin(2\pi y)+1$ for $y=0$ to 1. The quantum simulations use a second-order one-sided upwind finite difference stencil to evaluate the matrix $A$, where the direction of the stencil is in the opposite direction to the local velocity component. The maximum CFL parameter $r_\text{max} = 0.1$, the Hamiltonian evolution time per time step $\theta = \pi/2$, and the simulations have been carried out for $T=2800$ time steps with each time step occurring successfully. The initial conditions and subsequent evolution of the scalar field is shown in Fig.\ \ref{fig:lid_driven_cavity}b, which shows the scalar field swirling and being distorted by the velocity field. The no-slip wall naturally leads to a Dirichlet condition at the boundary in pure advection problems, and Fig.\ \ref{fig:lid_driven_cavity}b shows that the wall boundary values are effectively maintained. The solution is physically plausible and captures other qualitative features of the flow such as the vortex location, and is in close quantitative agreement with corresponding classical simulations.

\subsection{Effects of Noise and the Spatial Scheme}

Returning to the laminar channel flow where an analytical solution is available, Fig.\ \ref{fig:noise_contours} shows the effects of various types of noise on the simulations using $r_\text{max} = 0.1$, $\theta = \pi/2$ and a fourth-order central scheme for the spatial derivatives. Figure \ref{fig:noise_contours}a shows that, when subject to Gaussian noise with a standard deviation of 10\% of $\text{mean}(\phi)$, the numerical methods in the algorithm are capable of handling the noise as the same qualitative solution is obtained as the noise-free simulations in Fig.\ \ref{fig:advection_evolution}. The initial noise is retained in the final solution and does not appear to accumulate or dampen. Figure \ref{fig:noise_contours}b demonstrates the effects of noise in the Hamiltonian embedding procedure in Eq.\ \eqref{eq:hamiltonian_embedding}. When the entries of $A$ are subject to noise with a standard deviation of 1\% of the true value, the correct qualitative solution is obtained with errors that appear to grow linearly with time and the local value of $u(y)$.

\begin{figure}[tb]
    \centering
    \includegraphics[width=\columnwidth]{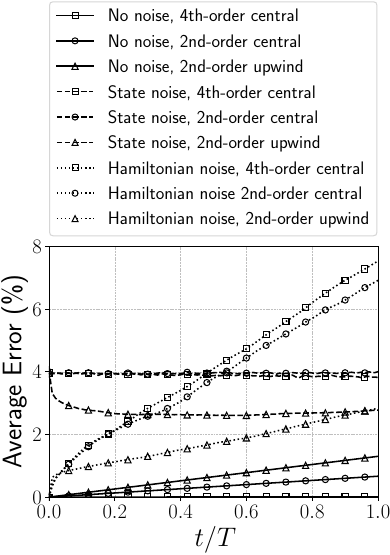}
    \caption{Mean absolute value of the error with different noise and spatial discretisation schemes.}
    \label{fig:noise_error}
\end{figure}

The growth of the mean absolute value of the error with no noise, initial state noise and Hamiltonian embedding noise is shown in Fig.\ \ref{fig:noise_error} for a fourth-order central scheme, a second-order central scheme and a second-order one-sided upwind scheme. In all cases, the errors are calculated against the noise-free analytical solution in Eq.\ \eqref{eq:analytical_solution}. Starting with the noise-free error evolution, Fig.\ \ref{fig:noise_error} confirms that the mean error grows linearly with time for all spatial discretisation schemes. The second-order one-sided scheme is the least accurate, with mean errors reaching 1.3\% after 2000 time steps, followed by the second-order and fourth-order central schemes with errors reaching 0.6\% and 0.1\% respectively. In the presence of noise, the one-sided scheme performs best due to the greater numerical dissipation of the noise in the small scales whilst still capturing the derivative information in the large scales. For the initial state noise, the central schemes maintain the mean error at approximately 4\% while the one-sided scheme sharply reduces the error before it stabilises at below 3\%. In the case of Hamiltonian embedding noise, errors grow linearly with the simulation time but with a much steeper gradient than the noise-free simulations. Again, the one-sided scheme significantly outperforms as the errors grow at a reduced rate.

\section{Conclusions}
\label{sec:conclusions}

A quantum algorithm for solving the advection equation, a linear PDE prevalent in various scientific and engineering industries, was presented. The algorithm uses sparse Hamiltonian simulation to embed the discrete time-marching operator $A$ into the Hamiltonian $H$, resulting in a unitary operator $\Omega=e^{-iH\theta}$ that encodes $A$ to a high accuracy regardless of the values of $\theta$. Postselection failure does not require further state initialisation queries since the resulting operation closely approximates the identity matrix, having a minimal impact on the quantum state and allowing the computation to continue. The algorithm applies to multidimensional problems with arbitrary boundary conditions and finite difference stencils.

From a resource utilisation perspective, qubit requirements grow logarithmically with the number of grid points $N$ and the circuit depth grows linearly with the desired number of time steps, the sparsity of the discrete time-marching operator and the inverse of the desired error, $\widetilde{O} (N_T s/\epsilon)$, when suppressing poly-logarithmic terms. This represents a significant polynomial speedup in complexity compared to classical methods, which typically exhibit a scaling of $O(N N_T)$, leading to a polynomial improvement by a factor of $N$.

It was demonstrated mathematically that the methodology does not universally apply to all PDEs, using the heat equation as an example. The derived mathematical expressions for the effects of the $r$ and $\theta$ parameters on the error have been validated numerically with statevector simulations. In the channel flow configuration, the amplitudes closely agreed with the analytical solution for all parameters tested. It was shown that there is an advantage in using high-order spatial schemes when the state varies as a continuous function, but that low-order dissipative schemes may outperform in noisy environments.

The typical advantages of the proposed approach over other algorithms are that the runtime grows linearly with the simulation time, the state can be reused on postselection failure requiring a single copy of the initial state, any combination of finite difference stencils can be used, and for its simplicity of implementation. Other algorithms that can be applied to the advection equation are typically aimed at solving homogeneous ODEs in the form of Eq.\ \eqref{eq:ODE}, which the advection equation reduces to when discretised in space. In comparison to QLSA-based algorithms \citep{Berry2014, Berry2017, Berry2022, Krovi2022}, the proposed approach excels due to the single copy of the initial quantum state required. The optimal QLSA \cite{Costa2022} requires $O(\kappa \log(1/\epsilon))$ queries to the state preparation oracle, which can become prohibitive when pursuing a practical quantum advantage. The quantum time-marching algorithm \cite{Fang2023} is conceptually the most similar to the algorithm presented here, although results in a runtime with a quadratic dependence on $T$ compared to a linear dependence in the present algorithm. Schr\"{o}dingerisation \cite{Jin2022, Jin2023, Jin2024} and LCHS \cite{An2023PhysRevLett, An2023} algorithms both assume that the Hermitian part of the coefficient matrix $M$ is negative semi-definite, which applies to central and upwind finite difference schemes for the advection equation. Occasionally, however, the use of downwind schemes cannot be avoided, such as when resolving the flow near a computational boundary. Therefore, the flexibility of the proposed approach and its simplicity of implementation make it a preferred choice.

Finally, developing algorithms that evolve a quantum state by the PDE of interest is only a step towards achieving a practical quantum advantage, with methods for efficiently preparing the state and extracting useful global statistics being crucial for preparing scientific and engineering industries to be quantum-ready.

\begin{acknowledgments}
  This research was funded by the Engineering and Physical Sciences Research Council in the United Kingdom, grant number EP/W032643/1.
\end{acknowledgments}

\bibliography{apssamp}

\begin{thebibliography}{49}%
\makeatletter
\providecommand \@ifxundefined [1]{%
 \@ifx{#1\undefined}
}%
\providecommand \@ifnum [1]{%
 \ifnum #1\expandafter \@firstoftwo
 \else \expandafter \@secondoftwo
 \fi
}%
\providecommand \@ifx [1]{%
 \ifx #1\expandafter \@firstoftwo
 \else \expandafter \@secondoftwo
 \fi
}%
\providecommand \natexlab [1]{#1}%
\providecommand \enquote  [1]{``#1''}%
\providecommand \bibnamefont  [1]{#1}%
\providecommand \bibfnamefont [1]{#1}%
\providecommand \citenamefont [1]{#1}%
\providecommand \href@noop [0]{\@secondoftwo}%
\providecommand \href [0]{\begingroup \@sanitize@url \@href}%
\providecommand \@href[1]{\@@startlink{#1}\@@href}%
\providecommand \@@href[1]{\endgroup#1\@@endlink}%
\providecommand \@sanitize@url [0]{\catcode `\\12\catcode `\$12\catcode
  `\&12\catcode `\#12\catcode `\^12\catcode `\_12\catcode `\%12\relax}%
\providecommand \@@startlink[1]{}%
\providecommand \@@endlink[0]{}%
\providecommand \url  [0]{\begingroup\@sanitize@url \@url }%
\providecommand \@url [1]{\endgroup\@href {#1}{\urlprefix }}%
\providecommand \urlprefix  [0]{URL }%
\providecommand \Eprint [0]{\href }%
\providecommand \doibase [0]{https://doi.org/}%
\providecommand \selectlanguage [0]{\@gobble}%
\providecommand \bibinfo  [0]{\@secondoftwo}%
\providecommand \bibfield  [0]{\@secondoftwo}%
\providecommand \translation [1]{[#1]}%
\providecommand \BibitemOpen [0]{}%
\providecommand \bibitemStop [0]{}%
\providecommand \bibitemNoStop [0]{.\EOS\space}%
\providecommand \EOS [0]{\spacefactor3000\relax}%
\providecommand \BibitemShut  [1]{\csname bibitem#1\endcsname}%
\let\auto@bib@innerbib\@empty
\bibitem [{\citenamefont {Bravyi}\ \emph {et~al.}(2022)\citenamefont {Bravyi},
  \citenamefont {Dial}, \citenamefont {Gambetta}, \citenamefont {Gil},\ and\
  \citenamefont {Nazario}}]{Bravyi2022}%
  \BibitemOpen
  \bibfield  {author} {\bibinfo {author} {\bibfnamefont {S.}~\bibnamefont
  {Bravyi}}, \bibinfo {author} {\bibfnamefont {O.}~\bibnamefont {Dial}},
  \bibinfo {author} {\bibfnamefont {J.~M.}\ \bibnamefont {Gambetta}}, \bibinfo
  {author} {\bibfnamefont {D.}~\bibnamefont {Gil}},\ and\ \bibinfo {author}
  {\bibfnamefont {Z.}~\bibnamefont {Nazario}},\ }\bibfield  {title} {\bibinfo
  {title} {{The future of quantum computing with superconducting qubits}},\
  }\href@noop {} {\bibfield  {journal} {\bibinfo  {journal} {Journal of Applied
  Physics}\ }\textbf {\bibinfo {volume} {132}} (\bibinfo {year} {2022})},\
  \bibinfo {note} {160902}\BibitemShut {NoStop}%
\bibitem [{\citenamefont {Stephenson}\ \emph {et~al.}(2020)\citenamefont
  {Stephenson}, \citenamefont {Nadlinger}, \citenamefont {Nichol},
  \citenamefont {An}, \citenamefont {Drmota}, \citenamefont {Ballance},
  \citenamefont {Thirumalai}, \citenamefont {Goodwin}, \citenamefont {Lucas},\
  and\ \citenamefont {Ballance}}]{Stephenson2022}%
  \BibitemOpen
  \bibfield  {author} {\bibinfo {author} {\bibfnamefont {L.~J.}\ \bibnamefont
  {Stephenson}}, \bibinfo {author} {\bibfnamefont {D.~P.}\ \bibnamefont
  {Nadlinger}}, \bibinfo {author} {\bibfnamefont {B.~C.}\ \bibnamefont
  {Nichol}}, \bibinfo {author} {\bibfnamefont {S.}~\bibnamefont {An}}, \bibinfo
  {author} {\bibfnamefont {P.}~\bibnamefont {Drmota}}, \bibinfo {author}
  {\bibfnamefont {T.~G.}\ \bibnamefont {Ballance}}, \bibinfo {author}
  {\bibfnamefont {K.}~\bibnamefont {Thirumalai}}, \bibinfo {author}
  {\bibfnamefont {J.~F.}\ \bibnamefont {Goodwin}}, \bibinfo {author}
  {\bibfnamefont {D.~M.}\ \bibnamefont {Lucas}},\ and\ \bibinfo {author}
  {\bibfnamefont {C.~J.}\ \bibnamefont {Ballance}},\ }\bibfield  {title}
  {\bibinfo {title} {High-rate, high-fidelity entanglement of qubits across an
  elementary quantum network},\ }\href@noop {} {\bibfield  {journal} {\bibinfo
  {journal} {Physical Review Letters}\ }\textbf {\bibinfo {volume} {124}}
  (\bibinfo {year} {2020})},\ \bibinfo {note} {110501}\BibitemShut {NoStop}%
\bibitem [{\citenamefont {Webb}\ \emph {et~al.}(1998)\citenamefont {Webb},
  \citenamefont {De~Cuevas},\ and\ \citenamefont {Richmond}}]{Webb1998}%
  \BibitemOpen
  \bibfield  {author} {\bibinfo {author} {\bibfnamefont {D.~J.}\ \bibnamefont
  {Webb}}, \bibinfo {author} {\bibfnamefont {B.~A.}\ \bibnamefont
  {De~Cuevas}},\ and\ \bibinfo {author} {\bibfnamefont {C.~S.}\ \bibnamefont
  {Richmond}},\ }\bibfield  {title} {\bibinfo {title} {Improved advection
  schemes for ocean models},\ }\href@noop {} {\bibfield  {journal} {\bibinfo
  {journal} {Journal of Atmospheric and Oceanic Technology}\ }\textbf {\bibinfo
  {volume} {15}},\ \bibinfo {pages} {1171} (\bibinfo {year}
  {1998})}\BibitemShut {NoStop}%
\bibitem [{\citenamefont {Rood}(1987)}]{Rood1987}%
  \BibitemOpen
  \bibfield  {author} {\bibinfo {author} {\bibfnamefont {R.~B.}\ \bibnamefont
  {Rood}},\ }\bibfield  {title} {\bibinfo {title} {Numerical advection
  algorithms and their role in atmospheric transport and chemistry models},\
  }\href@noop {} {\bibfield  {journal} {\bibinfo  {journal} {Reviews of
  geophysics}\ }\textbf {\bibinfo {volume} {25}},\ \bibinfo {pages} {71}
  (\bibinfo {year} {1987})}\BibitemShut {NoStop}%
\bibitem [{\citenamefont {Samuel}\ and\ \citenamefont
  {Evonuk}(2010)}]{Samuel2010}%
  \BibitemOpen
  \bibfield  {author} {\bibinfo {author} {\bibfnamefont {H.}~\bibnamefont
  {Samuel}}\ and\ \bibinfo {author} {\bibfnamefont {M.}~\bibnamefont
  {Evonuk}},\ }\bibfield  {title} {\bibinfo {title} {Modeling advection in
  geophysical flows with particle level sets},\ }\href@noop {} {\bibfield
  {journal} {\bibinfo  {journal} {Geochemistry, Geophysics, Geosystems}\
  }\textbf {\bibinfo {volume} {11}} (\bibinfo {year} {2010})}\BibitemShut
  {NoStop}%
\bibitem [{\citenamefont {Bazilevs}\ \emph {et~al.}(2007)\citenamefont
  {Bazilevs}, \citenamefont {Calo}, \citenamefont {Tezduyar},\ and\
  \citenamefont {Hughes}}]{Bazilevs2007}%
  \BibitemOpen
  \bibfield  {author} {\bibinfo {author} {\bibfnamefont {Y.}~\bibnamefont
  {Bazilevs}}, \bibinfo {author} {\bibfnamefont {V.~M.}\ \bibnamefont {Calo}},
  \bibinfo {author} {\bibfnamefont {T.~E.}\ \bibnamefont {Tezduyar}},\ and\
  \bibinfo {author} {\bibfnamefont {T.~J.}\ \bibnamefont {Hughes}},\ }\bibfield
   {title} {\bibinfo {title} {Yz$\beta$ discontinuity capturing for
  advection-dominated processes with application to arterial drug delivery},\
  }\href@noop {} {\bibfield  {journal} {\bibinfo  {journal} {International
  Journal for Numerical Methods in Fluids}\ }\textbf {\bibinfo {volume} {54}},\
  \bibinfo {pages} {593} (\bibinfo {year} {2007})}\BibitemShut {NoStop}%
\bibitem [{\citenamefont {Diao}\ \emph {et~al.}(2004)\citenamefont {Diao},
  \citenamefont {Li},\ and\ \citenamefont {Fang}}]{Diao2004}%
  \BibitemOpen
  \bibfield  {author} {\bibinfo {author} {\bibfnamefont {N.}~\bibnamefont
  {Diao}}, \bibinfo {author} {\bibfnamefont {Q.}~\bibnamefont {Li}},\ and\
  \bibinfo {author} {\bibfnamefont {Z.}~\bibnamefont {Fang}},\ }\bibfield
  {title} {\bibinfo {title} {Heat transfer in ground heat exchangers with
  groundwater advection},\ }\href@noop {} {\bibfield  {journal} {\bibinfo
  {journal} {International Journal of Thermal Sciences}\ }\textbf {\bibinfo
  {volume} {43}},\ \bibinfo {pages} {1203} (\bibinfo {year}
  {2004})}\BibitemShut {NoStop}%
\bibitem [{\citenamefont {Feynman}(1982)}]{Feynman1982}%
  \BibitemOpen
  \bibfield  {author} {\bibinfo {author} {\bibfnamefont {R.~P.}\ \bibnamefont
  {Feynman}},\ }\bibfield  {title} {\bibinfo {title} {Simulating physics with
  computers},\ }\href@noop {} {\bibfield  {journal} {\bibinfo  {journal}
  {nternational Journal of Theoretical Physics}\ }\textbf {\bibinfo {volume}
  {21}} (\bibinfo {year} {1982})}\BibitemShut {NoStop}%
\bibitem [{\citenamefont {An}\ \emph {et~al.}(2022)\citenamefont {An},
  \citenamefont {Liu}, \citenamefont {Wang},\ and\ \citenamefont
  {Zhao}}]{An2022}%
  \BibitemOpen
  \bibfield  {author} {\bibinfo {author} {\bibfnamefont {D.}~\bibnamefont
  {An}}, \bibinfo {author} {\bibfnamefont {J.-P.}\ \bibnamefont {Liu}},
  \bibinfo {author} {\bibfnamefont {D.}~\bibnamefont {Wang}},\ and\ \bibinfo
  {author} {\bibfnamefont {Q.}~\bibnamefont {Zhao}},\ }\bibfield  {title}
  {\bibinfo {title} {A theory of quantum differential equation solvers:
  limitations and fast-forwarding},\ }\href@noop {} {\bibfield  {journal}
  {\bibinfo  {journal} {arXiv preprint arXiv:2211.05246}\ } (\bibinfo {year}
  {2022})}\BibitemShut {NoStop}%
\bibitem [{\citenamefont {Cao}\ \emph {et~al.}(2013)\citenamefont {Cao},
  \citenamefont {Papageorgiou}, \citenamefont {Petras}, \citenamefont {Traub},\
  and\ \citenamefont {Kais}}]{Cao2013}%
  \BibitemOpen
  \bibfield  {author} {\bibinfo {author} {\bibfnamefont {Y.}~\bibnamefont
  {Cao}}, \bibinfo {author} {\bibfnamefont {A.}~\bibnamefont {Papageorgiou}},
  \bibinfo {author} {\bibfnamefont {I.}~\bibnamefont {Petras}}, \bibinfo
  {author} {\bibfnamefont {J.}~\bibnamefont {Traub}},\ and\ \bibinfo {author}
  {\bibfnamefont {S.}~\bibnamefont {Kais}},\ }\bibfield  {title} {\bibinfo
  {title} {Quantum algorithm and circuit design solving the {Poisson}
  equation},\ }\href@noop {} {\bibfield  {journal} {\bibinfo  {journal} {New
  Journal of Physics}\ }\textbf {\bibinfo {volume} {15}} (\bibinfo {year}
  {2013})},\ \bibinfo {note} {013021}\BibitemShut {NoStop}%
\bibitem [{\citenamefont {Costa}\ \emph {et~al.}(2019)\citenamefont {Costa},
  \citenamefont {Jordan},\ and\ \citenamefont {Ostrander}}]{Costa2019}%
  \BibitemOpen
  \bibfield  {author} {\bibinfo {author} {\bibfnamefont {P.~C.}\ \bibnamefont
  {Costa}}, \bibinfo {author} {\bibfnamefont {S.}~\bibnamefont {Jordan}},\ and\
  \bibinfo {author} {\bibfnamefont {A.}~\bibnamefont {Ostrander}},\ }\bibfield
  {title} {\bibinfo {title} {Quantum algorithm for simulating the wave
  equation},\ }\href@noop {} {\bibfield  {journal} {\bibinfo  {journal}
  {Physical Review A}\ }\textbf {\bibinfo {volume} {99}} (\bibinfo {year}
  {2019})},\ \bibinfo {note} {012323}\BibitemShut {NoStop}%
\bibitem [{\citenamefont {Wang}\ \emph {et~al.}(2020)\citenamefont {Wang},
  \citenamefont {Wang}, \citenamefont {Li}, \citenamefont {Fan}, \citenamefont
  {Wei},\ and\ \citenamefont {Gu}}]{Wang2020}%
  \BibitemOpen
  \bibfield  {author} {\bibinfo {author} {\bibfnamefont {S.}~\bibnamefont
  {Wang}}, \bibinfo {author} {\bibfnamefont {Z.}~\bibnamefont {Wang}}, \bibinfo
  {author} {\bibfnamefont {W.}~\bibnamefont {Li}}, \bibinfo {author}
  {\bibfnamefont {L.}~\bibnamefont {Fan}}, \bibinfo {author} {\bibfnamefont
  {Z.}~\bibnamefont {Wei}},\ and\ \bibinfo {author} {\bibfnamefont
  {Y.}~\bibnamefont {Gu}},\ }\bibfield  {title} {\bibinfo {title} {Quantum fast
  {Poisson} solver: the algorithm and complete and modular circuit design},\
  }\href@noop {} {\bibfield  {journal} {\bibinfo  {journal} {Quantum
  Information Processing}\ }\textbf {\bibinfo {volume} {19}},\ \bibinfo {pages}
  {1} (\bibinfo {year} {2020})}\BibitemShut {NoStop}%
\bibitem [{\citenamefont {Childs}\ \emph {et~al.}(2021)\citenamefont {Childs},
  \citenamefont {Liu},\ and\ \citenamefont {Ostrander}}]{Childs2021}%
  \BibitemOpen
  \bibfield  {author} {\bibinfo {author} {\bibfnamefont {A.~M.}\ \bibnamefont
  {Childs}}, \bibinfo {author} {\bibfnamefont {J.-P.}\ \bibnamefont {Liu}},\
  and\ \bibinfo {author} {\bibfnamefont {A.}~\bibnamefont {Ostrander}},\
  }\bibfield  {title} {\bibinfo {title} {High-precision quantum algorithms for
  partial differential equations},\ }\href@noop {} {\bibfield  {journal}
  {\bibinfo  {journal} {Quantum}\ }\textbf {\bibinfo {volume} {5}} (\bibinfo
  {year} {2021})},\ \bibinfo {note} {574}\BibitemShut {NoStop}%
\bibitem [{\citenamefont {Clader}\ \emph {et~al.}(2013)\citenamefont {Clader},
  \citenamefont {Jacobs},\ and\ \citenamefont {Sprouse}}]{Clader2013}%
  \BibitemOpen
  \bibfield  {author} {\bibinfo {author} {\bibfnamefont {B.~D.}\ \bibnamefont
  {Clader}}, \bibinfo {author} {\bibfnamefont {B.~C.}\ \bibnamefont {Jacobs}},\
  and\ \bibinfo {author} {\bibfnamefont {C.~R.}\ \bibnamefont {Sprouse}},\
  }\bibfield  {title} {\bibinfo {title} {Preconditioned quantum linear system
  algorithm},\ }\href@noop {} {\bibfield  {journal} {\bibinfo  {journal}
  {Physical Review Letters}\ }\textbf {\bibinfo {volume} {110}} (\bibinfo
  {year} {2013})},\ \bibinfo {note} {250504}\BibitemShut {NoStop}%
\bibitem [{\citenamefont {Montanaro}\ and\ \citenamefont
  {Pallister}(2016)}]{Montanaro2016}%
  \BibitemOpen
  \bibfield  {author} {\bibinfo {author} {\bibfnamefont {A.}~\bibnamefont
  {Montanaro}}\ and\ \bibinfo {author} {\bibfnamefont {S.}~\bibnamefont
  {Pallister}},\ }\bibfield  {title} {\bibinfo {title} {Quantum algorithms and
  the finite element method},\ }\href@noop {} {\bibfield  {journal} {\bibinfo
  {journal} {Physical Review A}\ }\textbf {\bibinfo {volume} {93}} (\bibinfo
  {year} {2016})},\ \bibinfo {note} {032324}\BibitemShut {NoStop}%
\bibitem [{\citenamefont {Childs}\ and\ \citenamefont
  {Liu}(2020)}]{Childs2020}%
  \BibitemOpen
  \bibfield  {author} {\bibinfo {author} {\bibfnamefont {A.~M.}\ \bibnamefont
  {Childs}}\ and\ \bibinfo {author} {\bibfnamefont {J.-P.}\ \bibnamefont
  {Liu}},\ }\bibfield  {title} {\bibinfo {title} {Quantum spectral methods for
  differential equations},\ }\href@noop {} {\bibfield  {journal} {\bibinfo
  {journal} {Communications in Mathematical Physics}\ }\textbf {\bibinfo
  {volume} {375}},\ \bibinfo {pages} {1427} (\bibinfo {year}
  {2020})}\BibitemShut {NoStop}%
\bibitem [{\citenamefont {Lloyd}\ \emph {et~al.}(2020)\citenamefont {Lloyd},
  \citenamefont {De~Palma}, \citenamefont {Gokler}, \citenamefont {Kiani},
  \citenamefont {Liu}, \citenamefont {Marvian}, \citenamefont {Tennie},\ and\
  \citenamefont {Palmer}}]{Lloyd2020}%
  \BibitemOpen
  \bibfield  {author} {\bibinfo {author} {\bibfnamefont {S.}~\bibnamefont
  {Lloyd}}, \bibinfo {author} {\bibfnamefont {G.}~\bibnamefont {De~Palma}},
  \bibinfo {author} {\bibfnamefont {C.}~\bibnamefont {Gokler}}, \bibinfo
  {author} {\bibfnamefont {B.}~\bibnamefont {Kiani}}, \bibinfo {author}
  {\bibfnamefont {Z.-W.}\ \bibnamefont {Liu}}, \bibinfo {author} {\bibfnamefont
  {M.}~\bibnamefont {Marvian}}, \bibinfo {author} {\bibfnamefont
  {F.}~\bibnamefont {Tennie}},\ and\ \bibinfo {author} {\bibfnamefont
  {T.}~\bibnamefont {Palmer}},\ }\bibfield  {title} {\bibinfo {title} {Quantum
  algorithm for nonlinear differential equations},\ }\href@noop {} {\bibfield
  {journal} {\bibinfo  {journal} {arXiv preprint arxiv:2011.06571}\ } (\bibinfo
  {year} {2020})}\BibitemShut {NoStop}%
\bibitem [{\citenamefont {Liu}\ \emph {et~al.}(2021)\citenamefont {Liu},
  \citenamefont {Kolden}, \citenamefont {Krovi}, \citenamefont {Loureiro},
  \citenamefont {Trivisa},\ and\ \citenamefont {Childs}}]{Liu2021}%
  \BibitemOpen
  \bibfield  {author} {\bibinfo {author} {\bibfnamefont {J.-P.}\ \bibnamefont
  {Liu}}, \bibinfo {author} {\bibfnamefont {H.~{\O}.}\ \bibnamefont {Kolden}},
  \bibinfo {author} {\bibfnamefont {H.~K.}\ \bibnamefont {Krovi}}, \bibinfo
  {author} {\bibfnamefont {N.~F.}\ \bibnamefont {Loureiro}}, \bibinfo {author}
  {\bibfnamefont {K.}~\bibnamefont {Trivisa}},\ and\ \bibinfo {author}
  {\bibfnamefont {A.~M.}\ \bibnamefont {Childs}},\ }\bibfield  {title}
  {\bibinfo {title} {Efficient quantum algorithm for dissipative nonlinear
  differential equations},\ }\href@noop {} {\bibfield  {journal} {\bibinfo
  {journal} {Proceedings of the National Academy of Sciences}\ }\textbf
  {\bibinfo {volume} {118}},\ \bibinfo {pages} {e2026805118} (\bibinfo {year}
  {2021})}\BibitemShut {NoStop}%
\bibitem [{\citenamefont {Peruzzo}\ \emph {et~al.}(2014)\citenamefont
  {Peruzzo}, \citenamefont {McClean}, \citenamefont {Shadbolt}, \citenamefont
  {Yung}, \citenamefont {Zhou}, \citenamefont {Love}, \citenamefont
  {Aspuru-Guzik},\ and\ \citenamefont {O'brien}}]{Peruzzo2014}%
  \BibitemOpen
  \bibfield  {author} {\bibinfo {author} {\bibfnamefont {A.}~\bibnamefont
  {Peruzzo}}, \bibinfo {author} {\bibfnamefont {J.}~\bibnamefont {McClean}},
  \bibinfo {author} {\bibfnamefont {P.}~\bibnamefont {Shadbolt}}, \bibinfo
  {author} {\bibfnamefont {M.-H.}\ \bibnamefont {Yung}}, \bibinfo {author}
  {\bibfnamefont {X.-Q.}\ \bibnamefont {Zhou}}, \bibinfo {author}
  {\bibfnamefont {P.~J.}\ \bibnamefont {Love}}, \bibinfo {author}
  {\bibfnamefont {A.}~\bibnamefont {Aspuru-Guzik}},\ and\ \bibinfo {author}
  {\bibfnamefont {J.~L.}\ \bibnamefont {O'brien}},\ }\bibfield  {title}
  {\bibinfo {title} {A variational eigenvalue solver on a photonic quantum
  processor},\ }\href@noop {} {\bibfield  {journal} {\bibinfo  {journal}
  {Nature communications}\ }\textbf {\bibinfo {volume} {5}} (\bibinfo {year}
  {2014})},\ \bibinfo {note} {4213}\BibitemShut {NoStop}%
\bibitem [{\citenamefont {Lubasch}\ \emph {et~al.}(2020)\citenamefont
  {Lubasch}, \citenamefont {Joo}, \citenamefont {Moinier}, \citenamefont
  {Kiffner},\ and\ \citenamefont {Jaksch}}]{Lubasch2020}%
  \BibitemOpen
  \bibfield  {author} {\bibinfo {author} {\bibfnamefont {M.}~\bibnamefont
  {Lubasch}}, \bibinfo {author} {\bibfnamefont {J.}~\bibnamefont {Joo}},
  \bibinfo {author} {\bibfnamefont {P.}~\bibnamefont {Moinier}}, \bibinfo
  {author} {\bibfnamefont {M.}~\bibnamefont {Kiffner}},\ and\ \bibinfo {author}
  {\bibfnamefont {D.}~\bibnamefont {Jaksch}},\ }\bibfield  {title} {\bibinfo
  {title} {Variational quantum algorithms for nonlinear problems},\ }\href@noop
  {} {\bibfield  {journal} {\bibinfo  {journal} {Physical Review A}\ }\textbf
  {\bibinfo {volume} {101}} (\bibinfo {year} {2020})},\ \bibinfo {note}
  {010301}\BibitemShut {NoStop}%
\bibitem [{\citenamefont {Kyriienko}\ \emph {et~al.}(2021)\citenamefont
  {Kyriienko}, \citenamefont {Paine},\ and\ \citenamefont
  {Elfving}}]{Kyriienko2021}%
  \BibitemOpen
  \bibfield  {author} {\bibinfo {author} {\bibfnamefont {O.}~\bibnamefont
  {Kyriienko}}, \bibinfo {author} {\bibfnamefont {A.~E.}\ \bibnamefont
  {Paine}},\ and\ \bibinfo {author} {\bibfnamefont {V.~E.}\ \bibnamefont
  {Elfving}},\ }\bibfield  {title} {\bibinfo {title} {Solving nonlinear
  differential equations with differentiable quantum circuits},\ }\href@noop {}
  {\bibfield  {journal} {\bibinfo  {journal} {Physical Review A}\ }\textbf
  {\bibinfo {volume} {103}} (\bibinfo {year} {2021})},\ \bibinfo {note}
  {052416}\BibitemShut {NoStop}%
\bibitem [{\citenamefont {Jaksch}\ \emph {et~al.}(2023)\citenamefont {Jaksch},
  \citenamefont {Givi}, \citenamefont {Daley},\ and\ \citenamefont
  {Rung}}]{Jaksch2023}%
  \BibitemOpen
  \bibfield  {author} {\bibinfo {author} {\bibfnamefont {D.}~\bibnamefont
  {Jaksch}}, \bibinfo {author} {\bibfnamefont {P.}~\bibnamefont {Givi}},
  \bibinfo {author} {\bibfnamefont {A.~J.}\ \bibnamefont {Daley}},\ and\
  \bibinfo {author} {\bibfnamefont {T.}~\bibnamefont {Rung}},\ }\bibfield
  {title} {\bibinfo {title} {Variational quantum algorithms for computational
  fluid dynamics},\ }\href@noop {} {\bibfield  {journal} {\bibinfo  {journal}
  {AIAA Journal}\ }\textbf {\bibinfo {volume} {61}},\ \bibinfo {pages} {1885}
  (\bibinfo {year} {2023})}\BibitemShut {NoStop}%
\bibitem [{\citenamefont {Gourianov}\ \emph {et~al.}(2022)\citenamefont
  {Gourianov}, \citenamefont {Lubasch}, \citenamefont {Dolgov}, \citenamefont
  {van~den Berg}, \citenamefont {Babaee}, \citenamefont {Givi}, \citenamefont
  {Kiffner},\ and\ \citenamefont {Jaksch}}]{Gourianov2022}%
  \BibitemOpen
  \bibfield  {author} {\bibinfo {author} {\bibfnamefont {N.}~\bibnamefont
  {Gourianov}}, \bibinfo {author} {\bibfnamefont {M.}~\bibnamefont {Lubasch}},
  \bibinfo {author} {\bibfnamefont {S.}~\bibnamefont {Dolgov}}, \bibinfo
  {author} {\bibfnamefont {Q.~Y.}\ \bibnamefont {van~den Berg}}, \bibinfo
  {author} {\bibfnamefont {H.}~\bibnamefont {Babaee}}, \bibinfo {author}
  {\bibfnamefont {P.}~\bibnamefont {Givi}}, \bibinfo {author} {\bibfnamefont
  {M.}~\bibnamefont {Kiffner}},\ and\ \bibinfo {author} {\bibfnamefont
  {D.}~\bibnamefont {Jaksch}},\ }\bibfield  {title} {\bibinfo {title} {A
  quantum-inspired approach to exploit turbulence structures},\ }\href@noop {}
  {\bibfield  {journal} {\bibinfo  {journal} {Nature Computational Science}\
  }\textbf {\bibinfo {volume} {2}},\ \bibinfo {pages} {30} (\bibinfo {year}
  {2022})}\BibitemShut {NoStop}%
\bibitem [{\citenamefont {Berry}(2014)}]{Berry2014}%
  \BibitemOpen
  \bibfield  {author} {\bibinfo {author} {\bibfnamefont {D.~W.}\ \bibnamefont
  {Berry}},\ }\bibfield  {title} {\bibinfo {title} {High-order quantum
  algorithm for solving linear differential equations},\ }\href@noop {}
  {\bibfield  {journal} {\bibinfo  {journal} {Journal of Physics A:
  Mathematical and Theoretical}\ }\textbf {\bibinfo {volume} {47}} (\bibinfo
  {year} {2014})},\ \bibinfo {note} {105301}\BibitemShut {NoStop}%
\bibitem [{\citenamefont {Berry}\ \emph {et~al.}(2017)\citenamefont {Berry},
  \citenamefont {Childs}, \citenamefont {Ostrander},\ and\ \citenamefont
  {Wang}}]{Berry2017}%
  \BibitemOpen
  \bibfield  {author} {\bibinfo {author} {\bibfnamefont {D.~W.}\ \bibnamefont
  {Berry}}, \bibinfo {author} {\bibfnamefont {A.~M.}\ \bibnamefont {Childs}},
  \bibinfo {author} {\bibfnamefont {A.}~\bibnamefont {Ostrander}},\ and\
  \bibinfo {author} {\bibfnamefont {G.}~\bibnamefont {Wang}},\ }\bibfield
  {title} {\bibinfo {title} {Quantum algorithm for linear differential
  equations with exponentially improved dependence on precision},\ }\href@noop
  {} {\bibfield  {journal} {\bibinfo  {journal} {Communications in Mathematical
  Physics}\ }\textbf {\bibinfo {volume} {356}},\ \bibinfo {pages} {1057}
  (\bibinfo {year} {2017})}\BibitemShut {NoStop}%
\bibitem [{\citenamefont {Arrazola}\ \emph {et~al.}(2019)\citenamefont
  {Arrazola}, \citenamefont {Kalajdzievski}, \citenamefont {Weedbrook},\ and\
  \citenamefont {Lloyd}}]{Arrazola2019}%
  \BibitemOpen
  \bibfield  {author} {\bibinfo {author} {\bibfnamefont {J.~M.}\ \bibnamefont
  {Arrazola}}, \bibinfo {author} {\bibfnamefont {T.}~\bibnamefont
  {Kalajdzievski}}, \bibinfo {author} {\bibfnamefont {C.}~\bibnamefont
  {Weedbrook}},\ and\ \bibinfo {author} {\bibfnamefont {S.}~\bibnamefont
  {Lloyd}},\ }\bibfield  {title} {\bibinfo {title} {Quantum algorithm for
  nonhomogeneous linear partial differential equations},\ }\href@noop {}
  {\bibfield  {journal} {\bibinfo  {journal} {Physical Review A}\ }\textbf
  {\bibinfo {volume} {100}} (\bibinfo {year} {2019})},\ \bibinfo {note}
  {032306}\BibitemShut {NoStop}%
\bibitem [{\citenamefont {Krovi}(2022)}]{Krovi2022}%
  \BibitemOpen
  \bibfield  {author} {\bibinfo {author} {\bibfnamefont {H.}~\bibnamefont
  {Krovi}},\ }\bibfield  {title} {\bibinfo {title} {Improved quantum algorithms
  for linear and nonlinear differential equations},\ }\href@noop {} {\bibfield
  {journal} {\bibinfo  {journal} {arXiv preprint arXiv:2202.01054}\ } (\bibinfo
  {year} {2022})}\BibitemShut {NoStop}%
\bibitem [{\citenamefont {Berry}\ and\ \citenamefont
  {Costa}(2022)}]{Berry2022}%
  \BibitemOpen
  \bibfield  {author} {\bibinfo {author} {\bibfnamefont {D.~W.}\ \bibnamefont
  {Berry}}\ and\ \bibinfo {author} {\bibfnamefont {P.}~\bibnamefont {Costa}},\
  }\bibfield  {title} {\bibinfo {title} {Quantum algorithm for time-dependent
  differential equations using dyson series},\ }\href@noop {} {\bibfield
  {journal} {\bibinfo  {journal} {arXiv preprint arXiv:2212.03544}\ } (\bibinfo
  {year} {2022})}\BibitemShut {NoStop}%
\bibitem [{\citenamefont {Harrow}\ \emph {et~al.}(2009)\citenamefont {Harrow},
  \citenamefont {Hassidim},\ and\ \citenamefont {Lloyd}}]{Harrow2009}%
  \BibitemOpen
  \bibfield  {author} {\bibinfo {author} {\bibfnamefont {A.~W.}\ \bibnamefont
  {Harrow}}, \bibinfo {author} {\bibfnamefont {A.}~\bibnamefont {Hassidim}},\
  and\ \bibinfo {author} {\bibfnamefont {S.}~\bibnamefont {Lloyd}},\ }\bibfield
   {title} {\bibinfo {title} {Quantum algorithm for linear systems of
  equations},\ }\href@noop {} {\bibfield  {journal} {\bibinfo  {journal}
  {Physical Review Letters}\ }\textbf {\bibinfo {volume} {103}} (\bibinfo
  {year} {2009})},\ \bibinfo {note} {150502}\BibitemShut {NoStop}%
\bibitem [{\citenamefont {Childs}\ \emph {et~al.}(2017)\citenamefont {Childs},
  \citenamefont {Kothari},\ and\ \citenamefont {Somma}}]{Childs2017}%
  \BibitemOpen
  \bibfield  {author} {\bibinfo {author} {\bibfnamefont {A.~M.}\ \bibnamefont
  {Childs}}, \bibinfo {author} {\bibfnamefont {R.}~\bibnamefont {Kothari}},\
  and\ \bibinfo {author} {\bibfnamefont {R.~D.}\ \bibnamefont {Somma}},\
  }\bibfield  {title} {\bibinfo {title} {Quantum algorithm for systems of
  linear equations with exponentially improved dependence on precision},\
  }\href@noop {} {\bibfield  {journal} {\bibinfo  {journal} {SIAM Journal on
  Computing}\ }\textbf {\bibinfo {volume} {46}},\ \bibinfo {pages} {1920}
  (\bibinfo {year} {2017})}\BibitemShut {NoStop}%
\bibitem [{\citenamefont {Costa}\ \emph {et~al.}(2022)\citenamefont {Costa},
  \citenamefont {An}, \citenamefont {Sanders}, \citenamefont {Su},
  \citenamefont {Babbush},\ and\ \citenamefont {Berry}}]{Costa2022}%
  \BibitemOpen
  \bibfield  {author} {\bibinfo {author} {\bibfnamefont {P.~C.}\ \bibnamefont
  {Costa}}, \bibinfo {author} {\bibfnamefont {D.}~\bibnamefont {An}}, \bibinfo
  {author} {\bibfnamefont {Y.~R.}\ \bibnamefont {Sanders}}, \bibinfo {author}
  {\bibfnamefont {Y.}~\bibnamefont {Su}}, \bibinfo {author} {\bibfnamefont
  {R.}~\bibnamefont {Babbush}},\ and\ \bibinfo {author} {\bibfnamefont {D.~W.}\
  \bibnamefont {Berry}},\ }\bibfield  {title} {\bibinfo {title} {Optimal
  scaling quantum linear-systems solver via discrete adiabatic theorem},\
  }\href@noop {} {\bibfield  {journal} {\bibinfo  {journal} {PRX quantum}\
  }\textbf {\bibinfo {volume} {3}},\ \bibinfo {pages} {040303} (\bibinfo {year}
  {2022})}\BibitemShut {NoStop}%
\bibitem [{\citenamefont {An}\ \emph {et~al.}(2023{\natexlab{a}})\citenamefont
  {An}, \citenamefont {Childs},\ and\ \citenamefont {Lin}}]{An2023}%
  \BibitemOpen
  \bibfield  {author} {\bibinfo {author} {\bibfnamefont {D.}~\bibnamefont
  {An}}, \bibinfo {author} {\bibfnamefont {A.~M.}\ \bibnamefont {Childs}},\
  and\ \bibinfo {author} {\bibfnamefont {L.}~\bibnamefont {Lin}},\ }\bibfield
  {title} {\bibinfo {title} {Quantum algorithm for linear non-unitary dynamics
  with near-optimal dependence on all parameters},\ }\href@noop {} {\bibfield
  {journal} {\bibinfo  {journal} {arXiv preprint arXiv:2312.03916}\ } (\bibinfo
  {year} {2023}{\natexlab{a}})}\BibitemShut {NoStop}%
\bibitem [{\citenamefont {Suau}\ \emph {et~al.}(2021)\citenamefont {Suau},
  \citenamefont {Staffelbach},\ and\ \citenamefont {Calandra}}]{Suau2021}%
  \BibitemOpen
  \bibfield  {author} {\bibinfo {author} {\bibfnamefont {A.}~\bibnamefont
  {Suau}}, \bibinfo {author} {\bibfnamefont {G.}~\bibnamefont {Staffelbach}},\
  and\ \bibinfo {author} {\bibfnamefont {H.}~\bibnamefont {Calandra}},\
  }\bibfield  {title} {\bibinfo {title} {Practical quantum computing: Solving
  the wave equation using a quantum approach},\ }\href@noop {} {\bibfield
  {journal} {\bibinfo  {journal} {ACM Transactions on Quantum Computing}\
  }\textbf {\bibinfo {volume} {2}},\ \bibinfo {pages} {1} (\bibinfo {year}
  {2021})}\BibitemShut {NoStop}%
\bibitem [{\citenamefont {Budinski}(2021)}]{Budinski2021}%
  \BibitemOpen
  \bibfield  {author} {\bibinfo {author} {\bibfnamefont {L.}~\bibnamefont
  {Budinski}},\ }\bibfield  {title} {\bibinfo {title} {Quantum algorithm for
  the advection--diffusion equation simulated with the lattice boltzmann
  method},\ }\href@noop {} {\bibfield  {journal} {\bibinfo  {journal} {Quantum
  Information Processing}\ }\textbf {\bibinfo {volume} {20}} (\bibinfo {year}
  {2021})}\BibitemShut {NoStop}%
\bibitem [{\citenamefont {Jin}\ \emph {et~al.}(2022)\citenamefont {Jin},
  \citenamefont {Liu},\ and\ \citenamefont {Yu}}]{Jin2022}%
  \BibitemOpen
  \bibfield  {author} {\bibinfo {author} {\bibfnamefont {S.}~\bibnamefont
  {Jin}}, \bibinfo {author} {\bibfnamefont {N.}~\bibnamefont {Liu}},\ and\
  \bibinfo {author} {\bibfnamefont {Y.}~\bibnamefont {Yu}},\ }\bibfield
  {title} {\bibinfo {title} {Quantum simulation of partial differential
  equations via schrodingerisation (2022)},\ }\href@noop {} {\bibfield
  {journal} {\bibinfo  {journal} {arXiv preprint arxiv:2212.13969}\ } (\bibinfo
  {year} {2022})}\BibitemShut {NoStop}%
\bibitem [{\citenamefont {Jin}\ \emph {et~al.}(2023)\citenamefont {Jin},
  \citenamefont {Liu},\ and\ \citenamefont {Yu}}]{Jin2023}%
  \BibitemOpen
  \bibfield  {author} {\bibinfo {author} {\bibfnamefont {S.}~\bibnamefont
  {Jin}}, \bibinfo {author} {\bibfnamefont {N.}~\bibnamefont {Liu}},\ and\
  \bibinfo {author} {\bibfnamefont {Y.}~\bibnamefont {Yu}},\ }\bibfield
  {title} {\bibinfo {title} {Quantum simulation of partial differential
  equations: Applications and detailed analysis},\ }\href@noop {} {\bibfield
  {journal} {\bibinfo  {journal} {Physical Review A}\ }\textbf {\bibinfo
  {volume} {108}} (\bibinfo {year} {2023})},\ \bibinfo {note}
  {032603}\BibitemShut {NoStop}%
\bibitem [{\citenamefont {Jin}\ \emph {et~al.}(2024)\citenamefont {Jin},
  \citenamefont {Li}, \citenamefont {Liu},\ and\ \citenamefont {Yu}}]{Jin2024}%
  \BibitemOpen
  \bibfield  {author} {\bibinfo {author} {\bibfnamefont {S.}~\bibnamefont
  {Jin}}, \bibinfo {author} {\bibfnamefont {X.}~\bibnamefont {Li}}, \bibinfo
  {author} {\bibfnamefont {N.}~\bibnamefont {Liu}},\ and\ \bibinfo {author}
  {\bibfnamefont {Y.}~\bibnamefont {Yu}},\ }\bibfield  {title} {\bibinfo
  {title} {Quantum simulation for partial differential equations with physical
  boundary or interface conditions},\ }\href@noop {} {\bibfield  {journal}
  {\bibinfo  {journal} {Journal of Computational Physics}\ }\textbf {\bibinfo
  {volume} {498}} (\bibinfo {year} {2024})},\ \bibinfo {note}
  {112707}\BibitemShut {NoStop}%
\bibitem [{\citenamefont {An}\ \emph {et~al.}(2023{\natexlab{b}})\citenamefont
  {An}, \citenamefont {Liu},\ and\ \citenamefont {Lin}}]{An2023PhysRevLett}%
  \BibitemOpen
  \bibfield  {author} {\bibinfo {author} {\bibfnamefont {D.}~\bibnamefont
  {An}}, \bibinfo {author} {\bibfnamefont {J.-P.}\ \bibnamefont {Liu}},\ and\
  \bibinfo {author} {\bibfnamefont {L.}~\bibnamefont {Lin}},\ }\bibfield
  {title} {\bibinfo {title} {Linear combination of hamiltonian simulation for
  nonunitary dynamics with optimal state preparation cost},\ }\href
  {https://doi.org/10.1103/PhysRevLett.131.150603} {\bibfield  {journal}
  {\bibinfo  {journal} {Phys. Rev. Lett.}\ }\textbf {\bibinfo {volume} {131}},\
  \bibinfo {pages} {150603} (\bibinfo {year} {2023}{\natexlab{b}})}\BibitemShut
  {NoStop}%
\bibitem [{\citenamefont {Fang}\ \emph {et~al.}(2023)\citenamefont {Fang},
  \citenamefont {Lin},\ and\ \citenamefont {Tong}}]{Fang2023}%
  \BibitemOpen
  \bibfield  {author} {\bibinfo {author} {\bibfnamefont {D.}~\bibnamefont
  {Fang}}, \bibinfo {author} {\bibfnamefont {L.}~\bibnamefont {Lin}},\ and\
  \bibinfo {author} {\bibfnamefont {Y.}~\bibnamefont {Tong}},\ }\bibfield
  {title} {\bibinfo {title} {Time-marching based quantum solvers for
  time-dependent linear differential equations},\ }\href@noop {} {\bibfield
  {journal} {\bibinfo  {journal} {Quantum}\ }\textbf {\bibinfo {volume} {7}}
  (\bibinfo {year} {2023})}\BibitemShut {NoStop}%
\bibitem [{\citenamefont {Berry}\ \emph {et~al.}(2015)\citenamefont {Berry},
  \citenamefont {Childs},\ and\ \citenamefont {Kothari}}]{Berry2015}%
  \BibitemOpen
  \bibfield  {author} {\bibinfo {author} {\bibfnamefont {D.~W.}\ \bibnamefont
  {Berry}}, \bibinfo {author} {\bibfnamefont {A.~M.}\ \bibnamefont {Childs}},\
  and\ \bibinfo {author} {\bibfnamefont {R.}~\bibnamefont {Kothari}},\
  }\bibfield  {title} {\bibinfo {title} {Hamiltonian simulation with nearly
  optimal dependence on all parameters},\ }in\ \href@noop {} {\emph {\bibinfo
  {booktitle} {IEEE 56th Annual Symposium on Foundations of Computer
  Science}}}\ (\bibinfo {year} {2015})\ pp.\ \bibinfo {pages}
  {792--809}\BibitemShut {NoStop}%
\bibitem [{\citenamefont {Shao}(2018)}]{Shao2018}%
  \BibitemOpen
  \bibfield  {author} {\bibinfo {author} {\bibfnamefont {C.}~\bibnamefont
  {Shao}},\ }\bibfield  {title} {\bibinfo {title} {Quantum algorithms to matrix
  multiplication},\ }\href@noop {} {\bibfield  {journal} {\bibinfo  {journal}
  {arXiv preprint arXiv:1803.01601}\ } (\bibinfo {year} {2018})}\BibitemShut
  {NoStop}%
\bibitem [{\citenamefont {Mitchell}\ and\ \citenamefont
  {Griffiths}(1980)}]{Mitchell1980}%
  \BibitemOpen
  \bibfield  {author} {\bibinfo {author} {\bibfnamefont {A.~R.}\ \bibnamefont
  {Mitchell}}\ and\ \bibinfo {author} {\bibfnamefont {D.~F.}\ \bibnamefont
  {Griffiths}},\ }\bibfield  {title} {\bibinfo {title} {The finite difference
  method in partial differential equations},\ }\href@noop {} {\bibfield
  {journal} {\bibinfo  {journal} {A Wiley-Interscience Publication}\ }
  (\bibinfo {year} {1980})}\BibitemShut {NoStop}%
\bibitem [{\citenamefont {Courant}\ \emph {et~al.}(1928)\citenamefont
  {Courant}, \citenamefont {Friedrichs},\ and\ \citenamefont
  {Lewy}}]{Courant1928}%
  \BibitemOpen
  \bibfield  {author} {\bibinfo {author} {\bibfnamefont {R.}~\bibnamefont
  {Courant}}, \bibinfo {author} {\bibfnamefont {K.}~\bibnamefont
  {Friedrichs}},\ and\ \bibinfo {author} {\bibfnamefont {H.}~\bibnamefont
  {Lewy}},\ }\bibfield  {title} {\bibinfo {title} {{\"U}ber die partiellen
  differenzengleichungen der mathematischen physik},\ }\href@noop {} {\bibfield
   {journal} {\bibinfo  {journal} {Mathematische Annalen}\ }\textbf {\bibinfo
  {volume} {100}},\ \bibinfo {pages} {32} (\bibinfo {year} {1928})}\BibitemShut
  {NoStop}%
\bibitem [{\citenamefont {Gingrich}\ and\ \citenamefont
  {Williams}(2004)}]{Gingrich2004}%
  \BibitemOpen
  \bibfield  {author} {\bibinfo {author} {\bibfnamefont {R.~M.}\ \bibnamefont
  {Gingrich}}\ and\ \bibinfo {author} {\bibfnamefont {C.~P.}\ \bibnamefont
  {Williams}},\ }\bibfield  {title} {\bibinfo {title} {Non-unitary
  probabilistic quantum computing},\ }in\ \href@noop {} {\emph {\bibinfo
  {booktitle} {Proceedings of the Winter International Symposium on Information
  and Communication Technologies}}}\ (\bibinfo {year} {2004})\BibitemShut
  {NoStop}%
\bibitem [{\citenamefont {Childs}(2010)}]{Childs2010}%
  \BibitemOpen
  \bibfield  {author} {\bibinfo {author} {\bibfnamefont {A.~M.}\ \bibnamefont
  {Childs}},\ }\bibfield  {title} {\bibinfo {title} {On the relationship
  between continuous-and discrete-time quantum walk},\ }\href@noop {}
  {\bibfield  {journal} {\bibinfo  {journal} {Communications in Mathematical
  Physics}\ }\textbf {\bibinfo {volume} {294}},\ \bibinfo {pages} {581}
  (\bibinfo {year} {2010})}\BibitemShut {NoStop}%
\bibitem [{\citenamefont {Berry}\ and\ \citenamefont
  {Childs}(2009)}]{Berry2009}%
  \BibitemOpen
  \bibfield  {author} {\bibinfo {author} {\bibfnamefont {D.~W.}\ \bibnamefont
  {Berry}}\ and\ \bibinfo {author} {\bibfnamefont {A.~M.}\ \bibnamefont
  {Childs}},\ }\bibfield  {title} {\bibinfo {title} {Black-box hamiltonian
  simulation and unitary implementation},\ }\href@noop {} {\bibfield  {journal}
  {\bibinfo  {journal} {arXiv preprint arXiv:0910.4157}\ } (\bibinfo {year}
  {2009})}\BibitemShut {NoStop}%
\bibitem [{\citenamefont {Berry}\ \emph {et~al.}(2014)\citenamefont {Berry},
  \citenamefont {Childs}, \citenamefont {Cleve}, \citenamefont {Kothari},\ and\
  \citenamefont {Somma}}]{Berry2014exponential}%
  \BibitemOpen
  \bibfield  {author} {\bibinfo {author} {\bibfnamefont {D.~W.}\ \bibnamefont
  {Berry}}, \bibinfo {author} {\bibfnamefont {A.~M.}\ \bibnamefont {Childs}},
  \bibinfo {author} {\bibfnamefont {R.}~\bibnamefont {Cleve}}, \bibinfo
  {author} {\bibfnamefont {R.}~\bibnamefont {Kothari}},\ and\ \bibinfo {author}
  {\bibfnamefont {R.~D.}\ \bibnamefont {Somma}},\ }\bibfield  {title} {\bibinfo
  {title} {Exponential improvement in precision for simulating sparse
  hamiltonians},\ }in\ \href@noop {} {\emph {\bibinfo {booktitle} {Proceedings
  of the forty-sixth annual ACM symposium on Theory of computing}}}\ (\bibinfo
  {year} {2014})\ pp.\ \bibinfo {pages} {283--292}\BibitemShut {NoStop}%
\bibitem [{\citenamefont {Pope}(2000)}]{Pope2000}%
  \BibitemOpen
  \bibfield  {author} {\bibinfo {author} {\bibfnamefont {S.~B.}\ \bibnamefont
  {Pope}},\ }\href@noop {} {\emph {\bibinfo {title} {Turbulent flows}}}\
  (\bibinfo  {publisher} {Cambridge University Press},\ \bibinfo {year}
  {2000})\BibitemShut {NoStop}%
\bibitem [{Ans(2024)}]{Ansys2024}%
  \BibitemOpen
  \href@noop {} {\bibinfo {title} {{ANSYS Fluent}}},\ \bibinfo {howpublished}
  {ANSYS, Inc.} (\bibinfo {year} {2024}),\ \bibinfo {note} {software available
  from ANSYS, Inc.}\BibitemShut {Stop}%
\end{thebibliography}%

\end{document}